\documentclass[12pt,a4paper,final]{iopart}
\expandafter\let\csname equation*\endcsname\relax
\expandafter\let\csname endequation*\endcsname\relax
\usepackage{iopams}
\usepackage{graphicx}
\usepackage{subfigure}
\usepackage[breaklinks=true,colorlinks=true,linkcolor=blue,urlcolor=blue,citecolor=blue]{hyperref}
\usepackage{cite}
\usepackage{amsopn}
\usepackage{braket}
\usepackage{mathtools}

\usepackage[a4paper]{geometry}
\usepackage{bpchem,upgreek}
	\usepackage{amsmath}
	\usepackage{makeidx}
	\usepackage{amsfonts}
	\usepackage[ansinew]{inputenc}
	\usepackage[usenames,dvipsnames]{pstricks}
    \usepackage{graphicx}
    \usepackage{float}
    \usepackage{subfigure}
	\usepackage{pst-grad} 
	\usepackage{pst-plot} 
	\usepackage{sidecap}
	\usepackage[makeroom]{cancel}

\begin{document}

\title[Distributed entangled state...]{Distributed entangled state production by using quantum repeater protocol}

\author{M Ghasemi$^{1}$}
\author{M K Tavassoly$^{1}$}
\address{$^1$Atomic and Molecular Group, Faculty of Physics, Yazd University, Yazd  89195-741, Iran}
\ead{mktavassoly@yazd.ac.ir}

\vspace{10pt}
\date{today}

\begin{abstract}
We consider entangled state production utilizing a full optomechanical arrangement, based on which we create entanglement between two far three-level V-type atoms using a quantum repeater protocol. At first, we consider eight identical atoms $(1,2,\cdots, 8)$, while adjacent pairs $(i,i+1)$ with $i=1,3,5,7$ have been prepared in entangled states and
 the atoms 1, 8 are the two target atoms. The three-level atoms (1,2,3,4) and (5,6,7,8) distinctly become entangled with the system including optical and mechanical modes
 by performing the interaction in optomechanical cavities between atoms (2,3) and (6,7), respectively. Then, by operating appropriate measurements, instead of Bell state measurement which is a hard task in practical works, the entangled states of atoms (1,4) and (5,8) are achieved.
 Next, via interacting atoms (4,5) of the pairs (1,4) and (5,8) and operating proper measurement,
the entangled state of target atoms (1,8) is obtained.
 In the continuation, entropy and success probability of the produced entangled state are then evaluated. It is observed that the time period of entropy is increased by increasing the mechanical
 frequency ($\omega_M$) and by decreasing optomechanical coupling strength to the field modes ($G$). Also, in most cases, the
 maximum of success probability is increased by decreasing $G$
 and via decreasing $\omega_M$.
\end{abstract}


\vspace{2pc}
%
%
%
%

\section{Introduction}
Distribution of entangled photon states over long distances that plays an important role in quantum information is limited because of the existence of photonic absorption \cite{Briegel1998} via various processes. Quantum repeater is an invasion to this unwanted happening wherein by using this protocol, in principle,  the entanglement can be distributed over long distances \cite{Bernad2017,Ghasemi22019,Ghasemi32019}.  In this protocol, long distances are divided into several short parts, where in each part a bipartite entangled state is set, separable from the adjacent pairs. Then, the entanglement from the entangled pairs is swapped between the distinct (separable) parts with the help of the beam splitter \cite{Agarwal2012,Pakniat2017}, using Bell state measurement method \cite{Ghasemi2016,Ghasemi2017} or by performing the interaction \cite{Nourmandipour12016,Pakniat2016,Yang2005,Pakniat20172,Pakniat2018} described by Jaynes-Cummings model (JCM) \cite{Jaynes1963,Shen2017} or Tavis-Cummings model (TCM) \cite{Tavis1968} (for more information see \cite{Wang20112,Wang20132,Song20182}).
  Even though entanglement generation is indeed in the core of the quantum repeater protocols, entanglement swapping and quantum memory are the other two important elements of quantum repeater (quantum memory is needed for future enough long distance quantum communication \cite{Tittel2010}).
    In other words, after producing entanglement between segments of separate states, the produced entangled states store and release by quantum memories and without them, all probabilistic steps would have to succeed at the same time \cite{Tittel2010}.
   Nowadays, quantum repeater protocols have been investigated on quantum dots \cite{Li2016}, entangled photons \cite{Wang2012}, Rydberg gates \cite{Zhao2010} and dispersive cavity quantum electrodynamics interactions between matter qubits and bright coherent light \cite{Ladd2006}. Also, quantum repeater based on atomic state has been investigated in \cite{Yi2019} where the authors have produced distributed maximally entangled state by performing interaction on separable atoms in three optical cavities.\\
    In our previous works \cite{Ghasemi22019,Ghasemi32019,Ghasemi2018}, we have considered distributed atomic entangled states production based on quantum repeater protocol by performing interaction in optical cavities. In \cite{Ghasemi2019} we have investigated quantum repeater using coherent state by utilizing the beam splitter. Also, we have considered quantum repeater protocol for the QED-optomechanics hybrid system in \cite{Ghasemi42019}. In \cite{Ghasemi2020}, we designed the repeater protocol which is partly in the presence of optomechanical cavity (OMC) to distribute the entangled state of atomic ions. In the present paper, however, our goal is to distribute atomic entangled state in the presence of a full optomechanical arrangement system (see figure \ref{fig.Fig1a}). The dissipation effects generally exist in any quantum interacting systems and so in optomechanical systems. In this respect
     there exists literature in which the decoherence effects are considered \cite{Bougouffa2020,Rehaily2017,Bougouffa20202,Awfi2018}.
     Moreover, our system has been assumed to consist a high-Q cavity $(Q\sim 4\times 10^{10})$ in which the optical decoherences can be ignored \cite{Joshi2004,Weidinger1999}. Also, the mechanical damping
     and input noise can be neglected when the interaction time is shorter
     than the decoherence time scale \cite{Vanner2013,Li2018,Ho2018}. Accordingly, phonon- and photon- damping have been ignored. 
    We would like to mention our motivation at this point. In this regard, as is well-known preparing and detecting macroscopic quantum phenomena is of enough significance. Optomechanical systems have potential ability to provide explicit evidences on quantum features in macroscopically arrangement \cite{Schwab2005}.
     In cavity optomechanical systems, the optical cavity with one movable mirror is concerned with mechanical effects caused by light through radiation pressure \cite{Liu2013,Nadiki2016,Nadiki2017,Nadiki2018}. OMC includes a coupling between photons and phonons, so that OMCs are useful in quantum information processing in order to combine the advantages of different physical systems in a unit architecture \cite{Aspelmeyer2014}.
          Also, mechanical resonators are coupled to many different types of quantum two-level systems such as trapped atoms in an optical lattice \cite{Camerer2011}.
          Atomic systems play a key role in the interaction procedure in OMCs wherein entanglement generation of atomic systems is a good paradigm 
          \cite{Ian2008,Wang2018,Bai2016,Liao2016,Liao20162,Barzanjeh2011}. In \cite{Zhou2011}, the authors have investigated the entanglement between two-mode
           fields in an optomechanical system with two movable mirrors.
    There are several possible applications of optomechanical devices \cite{Aspelmeyer2014} such as an all-optical memory element \cite{Bagheri2011} and a new technology for single-photon detection \cite{Ludwig2012}.
               Therefore, introducing a setup consists of an arrangement of a number of OMCs may be of interest in quantum
          information processing like quantum repeater.\\
\\
      This paper organizes as follows: quantum repeater protocol for producing the distributed entangled state is introduced in Sec.
      \ref{model}. The numerical results and discussion are investigated in Sec. \ref{sec.results}.  Finally, the paper ends with a summary and
      conclusions in Sec. \ref{Summary}.
      \section{Quantum repeater protocol}\label{model}
              \begin{figure}[H]
       \centering
           \subfigure[\label{fig.Fig1a} \ The quantum repeater protocol]{\includegraphics[width=0.56\textwidth]{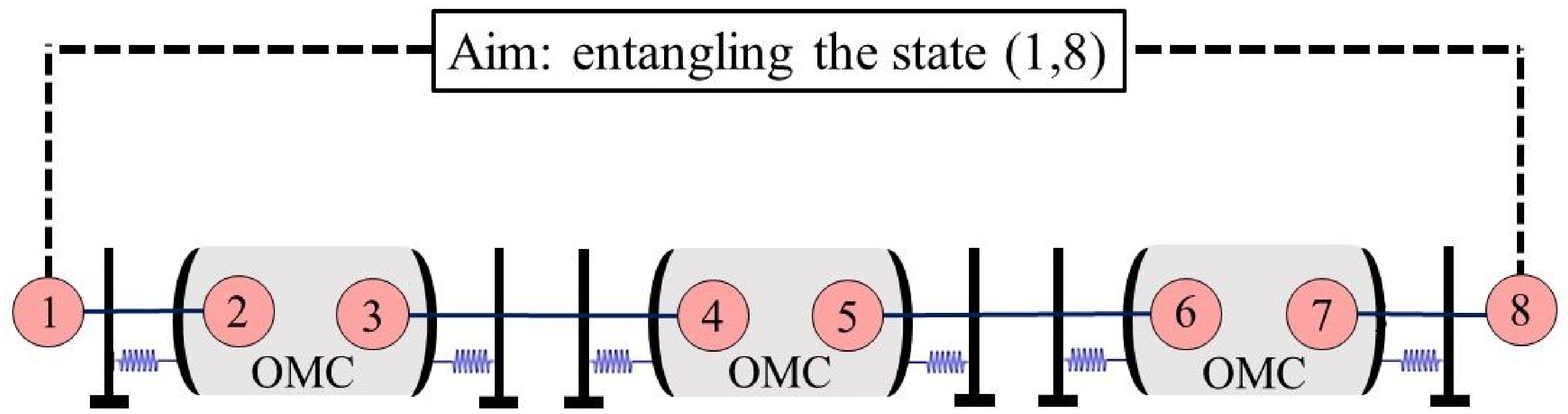}}
           \hspace{0.05\textwidth}
           \subfigure[\label{fig.Fig1b} \ The three-level V-type atoms]{\includegraphics[width=0.36\textwidth]{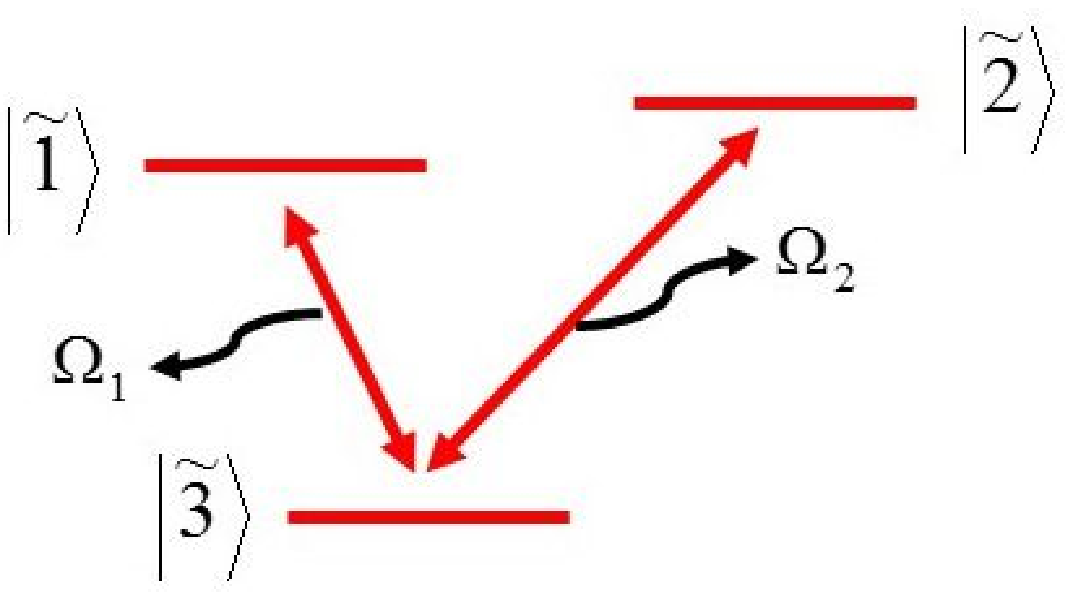}}
 
       \caption{\label{fig:Fig1} {(a) The scheme of quantum repeater protocol for distributed entangled state production using three OMCs. The atomic pairs (2,3), (6,7) and (4,5) are interacted in OMC, separately. The three OMCs are similar and in each cavity there are two identical atoms, two mechanical modes $(b_1, b^\dagger_1)$, $(b_2, b^\dagger_2)$ and two optical modes $(a_1, a^\dagger_1)$, $(a_2, a^\dagger_2)$. So, the interaction in each cavity can be introduced by a common Hamiltonian. (b) The diagram of three-level V-type atoms $(1,2,\cdots, 8)$. }      }
        \end{figure}
       In this section, we introduce quantum repeater protocol for producing the distributed entangled state between two far three-level V-type atoms 1 and 8 using a full optomechanical arrangement (see figure \ref{fig.Fig1a}). In fact, entanglement generation in the presence of OMCs is in general an attractive topic
           (see \cite{Zhou2011,Vitali2007,Wang2013,Hofer2011,Liao2014}). 
         It should be mentioned that there exist the effects of noises in entanglement generation in quantum systems \cite{Aloufi2015}. But note that, our system is considered in high-Q cavity and enough low temperature,
            so thermal noise can be sufficiently reduced. In this respect, it is established in  \cite{Courty2001} that thermal noise of mirrors in OMC can be reduced by cold damping. Also, a design for a novel optomechanical device capable of ultrahigh quality factors in the audio frequency band with negligible thermal noise has been reported in  \cite{Page2016}. In our protocol, eight three-level atoms $(1,2,\cdots, 8)$ are considered where the diagram of three-level atoms is shown in figure \ref{fig.Fig1b}. As is seen, the allowable transition-levels are shown. We suppose that the atomic pairs $(i,i+1)$ with $i=1,3,5,7$ have been initially prepared in the following entangled state:
            \begingroup\makeatletter\def\f@size{9}\check@mathfonts
          \begin{eqnarray}\label{initialstate}
       \ket{\psi(0)}_{i,i+1}&=&\frac{1}{\sqrt{2}}(\ket{\widetilde{1},\widetilde{3}}+\ket{\widetilde{3},\widetilde{1}})_{i,i+1}.
         \end{eqnarray}
       It should be mentioned that the atomic entangled states can be prepared experimentally by
       exciting two single $^{87}Rb$ atoms in remote traps
       and detecting interference of the emitted photons \cite{Rosenfeld2009}. Also, two three-level atoms can be entangled using interference of polarized photons \cite{Feng2003}. In \cite{Hofmann2012} the authors have demonstrated
       heralded entanglement between two atoms 20 m
       apart. In this study, at first, the spin of each of the two atoms is entangled with the polarization
       state of an emitted photon. Then, the photons are sent to a Bell state measurement setup and
       finally, the entanglement is generated between
       the atomic spins. The interaction between two V-type atoms (2,3) as well as (6,7) in the two independent OMCs is described by the
         Hamiltonian $\hat{H}_{(j,j+1)}=\hat{H}_0+\hat{H}_1$, with $j=2,6$ $(\hbar=1)$,
       \begin{eqnarray}\label{hamiltonian}
        \hat{H}_0&=&\sum^3_{i=1}\widetilde{\omega}_i\hat{\sigma}_{ii}^{(j)}+\sum^3_{i=1}\widetilde{\omega}_i\hat{\sigma}_{ii}^{(j+1)}+\sum_{i=1,2}\Omega_i\hat{a}_i^{\dagger}\hat{a}_i+\sum_{i=1,2}\omega_i\hat{b}_i^{\dagger}\hat{b}_i,\\\nonumber
        \hat{H}_1&=&\lambda_1(\hat{a}_1\hat{\sigma}_{13}^{(j)}+\hat{a}_1^{\dagger}\hat{\sigma}_{31}^{(j)})+\lambda_2(\hat{a}_2\hat{\sigma}_{23}^{(j)}+\hat{a}_2^{\dagger}\hat{\sigma}_{32}^{(j)})+\lambda^{'}_1(\hat{a}_1\hat{\sigma}_{13}^{(j+1)}+\hat{a}_1^{\dagger}\hat{\sigma}_{31}^{(j+1)})\\\nonumber
        &+&\lambda^{'}_2(\hat{a}_2\hat{\sigma}_{23}^{(j+1)}+\hat{a}_2^{\dagger}\hat{\sigma}_{32}^{(j+1)})-G\hat{a}_1^{\dagger}\hat{a}_1(\hat{b}_1+\hat{b}_1^{\dagger})-G^{'}\hat{a}_2^{\dagger}\hat{a}_2(\hat{b}_2+\hat{b}_2^{\dagger}).
         \end{eqnarray}
         The atomic transition operator has been shown by $\hat{\sigma}_{lm}^{(k)}=\ket{\widetilde{l}}_{(k)}\bra{\widetilde{m}}$ where $\widetilde{l},\widetilde{m}=\widetilde{1},\widetilde{2},\widetilde{3}$ denote the levels of atoms, $k=2,3,6,7$ refers to the atomic labels and $\widetilde{\omega}_i$ is atomic frequency. Also, $\hat{a}_1$, $\hat{a}_2$ ($\hat{b}_1$, $\hat{b}_2$) are two optical (mechanical) modes where the associated frequencies are denoted as $\Omega_1$, $\Omega_2$ ($\omega_1$, $\omega_2$), and $\lambda_i$, $\lambda^{'}_i$ ($G$, $G^{'}$) are the coupling constants between atoms (mechanical modes) with the optical modes.
       The Hamiltonian describing the interaction between atoms $(j,j+1)$ with $j=2,6$ in the interaction picture can be easily obtained as follows:
         \begin{eqnarray}\label{hbl}
          \hat{H}^\mathrm{int}_{(j,j+1)}&=&e^{i \hat{H}_0 t} \hat{H}_1e^{-i \hat{H}_0 t}\\ \nonumber
          &=&e^{i\omega_M t}\left[ \lambda_1\left( \hat{a}_1\hat{\sigma}^{(j)}_{13}+\hat{a}_1\hat{\sigma}^{(j+1)}_{13}\right)+\lambda_2\left( \hat{a}_2\hat{\sigma}^{(j)}_{23}+\hat{a}_2\hat{\sigma}^{(j+1)}_{23}\right)-G\left(  \hat{a}_1^{\dagger}\hat{a}_1\hat{b}^{\dagger}_1+ \hat{a}_2^{\dagger}\hat{a}_2\hat{b}^{\dagger}_2\right) \right] + h.c.,
              \end{eqnarray}
       where we assumed $\lambda_i=\lambda^{'}_i$, $i=1,2$, $G=G^{'}$, $\omega_1=\omega_2=\omega_M$ and $\widetilde{\omega}_1-\widetilde{\omega}_3-\Omega_1=\omega_M=\widetilde{\omega}_2-\widetilde{\omega}_3-\Omega_2$.
       With the help of the proposed approach in Refs. \cite{James2007,Gamel2010}, the effective Hamiltonian related to the interaction between atoms $(j,j+1)$ can be calculated, results in (the details are explained in Appendix A):
       \begin{eqnarray}\label{effectivehamiltonian}
       \hat{H}^{\mathrm{eff}}_{(j,j+1)}&=&\frac{\lambda^2_1}{\omega_M}\sum_{i=j,j+1}\left[ \hat{\sigma}_{11}^{(i)}+\hat{a}_1^{\dagger}\hat{a}_1\left( \hat{\sigma}_{11}^{(i)}-\hat{\sigma}_{33}^{(i)}\right) \right]+\frac{\lambda^2_2}{\omega_M}\sum_{i=j,j+1}\left[ \hat{\sigma}_{22}^{(i)}+\hat{a}_2^{\dagger}\hat{a}_2\left( \hat{\sigma}_{22}^{(i)}-\hat{\sigma}_{33}^{(i)}\right) \right] \\\nonumber
       &+&\frac{\lambda^2_1}{\omega_M}\left( \hat{\sigma}_{13}^{(j)} \hat{\sigma}_{31}^{(j+1)}+\hat{\sigma}_{31}^{(j)} \hat{\sigma}_{13}^{(j+1)}\right) +\frac{\lambda^2_2}{\omega_M}\left( \hat{\sigma}_{23}^{(j)} \hat{\sigma}_{32}^{(j+1)}+\hat{\sigma}_{32}^{(j)} \hat{\sigma}_{23}^{(j+1)}\right)-\frac{G^2}{\omega_M}\left[\left(\hat{a}_1^{\dagger}\hat{a}_1 \right)^2 +\left(\hat{a}_2^{\dagger}\hat{a}_2 \right)^2 \right] \\\nonumber
       & +&\frac{\lambda_1\lambda_2}{\omega_M}\left( \sum_{i=j,j+1}\hat{a}_1\hat{a}^{\dagger}_2  \hat{\sigma}_{12}^{(i)}+h.c.\right)-\frac{G \lambda_1}{\omega_M}\left( \sum_{i=j,j+1}\hat{a}_1\hat{b}_1  \hat{\sigma}_{13}^{(i)}+h.c.\right)-\frac{G \lambda_2}{\omega_M}\left( \sum_{i=j,j+1}\hat{a}_2\hat{b}_2  \hat{\sigma}_{23}^{(i)}+h.c.\right).
       \end{eqnarray}
        Using the effective Hamiltonian (\ref{effectivehamiltonian}) and paying our attention to the initial state $\ket{0,0;0,0} \otimes\ket{\psi(0)}_{1,2(5,6)}\otimes\ket{\psi(0)}_{3,4(7,8)}$, \textit{i.e.,} the optical and mechanical modes are all in vacuum states and the atoms are in the Bell-like state as in Eq. (\ref{initialstate}), the entangled state of atoms (1,2,3,4) (as well as atoms (5,6,7,8)), optical and mechanical modes at time $t$ can be stated as\footnote{Each ket of Eq. (\ref{state1-4}) is related to $\ket{(a_1, a^\dagger_1),(a_2, a^\dagger_2);(b_1, b^\dagger_1),(b_2, b^\dagger_2);\mathrm{atom}1(5),\mathrm{atom} 2(6);\mathrm{atom} 3(7),\mathrm{atom} 4(8)}$. For instance, the first term in state (\ref{state1-4}), \textit{i.e.,} $|0, 0; 0, 0;\widetilde{1},\widetilde{3};\widetilde{3},\widetilde{1} \rangle$, means that the optical and the mechanical
               modes are in the vacuum state $|0, 0 \rangle \otimes |0, 0 \rangle$ and the atoms 1, 2 (as well as 5, 6) are in the state
               $|\widetilde{1},\widetilde{3}\rangle$ and the atoms 3, 4 (as well as 7, 8) are in the state $|\widetilde{3},\widetilde{1}\rangle$.},
       \begin{eqnarray}\label{state1-4}
        \ket{\psi(t)}_{1-4(5-8)}&=&A_1(t)\ket{0,0;0,0;\widetilde{1},\widetilde{3};\widetilde{3},\widetilde{1}}+A_2(t)\ket{0,0;0,0;\widetilde{1},\widetilde{3};\widetilde{1},\widetilde{3}}\\\nonumber
         &+&A_3(t)\ket{0,0;0,0;\widetilde{1},\widetilde{1};\widetilde{3},\widetilde{3}}+A_4(t)\ket{1,0;1,0;\widetilde{1},\widetilde{3};\widetilde{3},\widetilde{3}}\\\nonumber
          &+&A_5(t)\ket{0,0;0,0;\widetilde{3},\widetilde{1};\widetilde{1},\widetilde{3}}+A_6(t)\ket{1,0;1,0;\widetilde{3},\widetilde{3};\widetilde{1},\widetilde{3}}\\\nonumber
         &+&A_7(t)\ket{1,0;1,0;\widetilde{3},\widetilde{1};\widetilde{3},\widetilde{3}}+A_8(t)\ket{2,0;2,0;\widetilde{3},\widetilde{3};\widetilde{3},\widetilde{3}}\\\nonumber
          &+&A_9(t)\ket{0,0;0,0;\widetilde{3},\widetilde{1};\widetilde{3},\widetilde{1}}+A_{10}(t)\ket{0,0;0,0;\widetilde{3},\widetilde{3};\widetilde{1},\widetilde{1}}\\\nonumber
           &+&A_{11}(t)\ket{1,0;1,0;\widetilde{3},\widetilde{3};\widetilde{3},\widetilde{1}}.
       \end{eqnarray}
     We explained how the above coefficients are obtained by us in Appendix B. Now, we apply the projection operators on state (\ref{state1-4}) which these operators are performed with the states $\ket{0,0} \otimes \ket{\widetilde{3},\widetilde{1}}$\footnote{The ket $|0, 0 \rangle$ (in the projection operator
                    with the state $|0, 0 \rangle \otimes |\widetilde{3},\widetilde{1}\rangle$) is related to optical and mechanical mode denoted by $(a_1, a^\dagger_1)$ and $(b_1, b^\dagger_1)$, respectively. Also, the state $|\widetilde{3},\widetilde{1}\rangle$ is related to atoms 2, 3 (and 6, 7). These explanations can be easily repeated for state $|0, 0 \rangle \otimes |\widetilde{1},\widetilde{3}\rangle$.} and $\ket{0,0} \otimes \ket{\widetilde{1},\widetilde{3}} $. As a result, after applying these operations, the atoms 1, 4 (as well as the atoms 5, 8) are respectively converted to the following entangled states,
       \begin{eqnarray}\label{state114}
           \ket{\psi(t)}^1_{1,4(5,8)}&=&\frac{1}{\sqrt{P(t)} }(A_2(t)\ket{\widetilde{1},\widetilde{3}}+A_{10}(t)\ket{\widetilde{3},\widetilde{1}}),
        \end{eqnarray}
        and
        \begin{eqnarray}\label{state214}
            \ket{\psi(t)}^2_{1,4(5,8)}&=&\frac{1}{\sqrt{P(t)} }(A_{10}(t)\ket{\widetilde{1},\widetilde{3}}+A_{2}(t)\ket{\widetilde{3},\widetilde{1}}),
         \end{eqnarray}
        (see Appendix B and notice that $A_2(t)=A_9(t)$, $A_3(t)=A_{10}(t)$). The entropy (showing the entanglement of atoms 1, 4 or atoms 5, 8) and success probability of arriving at the states (\ref{state114}), (\ref{state214}) have been obtained as
        \begin{eqnarray}\label{ent14}
            E(t)&=&1-\frac{\left| A_2(t)\right|^4+\left| A_{10}(t)\right|^4}{(P(t))^2},
          \end{eqnarray}
       \begin{eqnarray}\label{suc14}
          P(t)&=&\left| A_2(t)\right|^2+\left| A_{10}(t)\right|^2.
        \end{eqnarray}
        So, to perform the final interaction between atoms (4,5) in an OMC and achieve the entangled state for atoms (1,4,5,8) there exist four possible initial states $\ket{0,0;0,0}\otimes\ket{\psi(t)}^{1}_{1,4}\otimes\ket{\psi(t)}^{1}_{5,8}$, $\ket{0,0;0,0}\otimes\ket{\psi(t)}^{2}_{1,4}\otimes\ket{\psi(t)}^{2}_{5,8}$, $\ket{0,0;0,0}\otimes\ket{\psi(t)}^{1}_{1,4}\otimes\ket{\psi(t)}^{2}_{5,8}$ and $\ket{0,0;0,0}\otimes\ket{\psi(t)}^{2}_{1,4}\otimes\ket{\psi(t)}^{1}_{5,8}$. In the mentioned initial states, the states $\ket{\psi(t)}^{1}_{1,4(5,8)}$ and $\ket{\psi(t)}^{2}_{1,4(5,8)}$ have been defined in Eqs. (\ref{state114}) and (\ref{state214}), respectively. Also, we have assumed that the initial states for optical and mechanical modes in final interaction are all prepared in vacuum state $\ket{0,0;0,0}$. The interaction described by the following effective Hamiltonian,
        \begin{eqnarray}\label{effectivehamiltonian2}
               \hat{H}^{\mathrm{eff}}_{(4,5)}&=&\frac{\lambda^2_1}{\omega_M}\sum_{i=4,5}\left[ \hat{\sigma}_{11}^{(i)}+\hat{a}_1^{\dagger}\hat{a}_1\left( \hat{\sigma}_{11}^{(i)}-\hat{\sigma}_{33}^{(i)}\right) \right]+\frac{\lambda^2_2}{\omega_M}\sum_{i=4,5}\left[ \hat{\sigma}_{22}^{(i)}+\hat{a}_2^{\dagger}\hat{a}_2\left( \hat{\sigma}_{22}^{(i)}-\hat{\sigma}_{33}^{(i)}\right) \right] \\\nonumber
               &+&\frac{\lambda^2_1}{\omega_M}\left( \hat{\sigma}_{13}^{(4)} \hat{\sigma}_{31}^{(5)}+\hat{\sigma}_{31}^{(4)} \hat{\sigma}_{13}^{(5)}\right) +\frac{\lambda^2_2}{\omega_M}\left( \hat{\sigma}_{23}^{(4)} \hat{\sigma}_{32}^{(5)}+\hat{\sigma}_{32}^{(4)} \hat{\sigma}_{23}^{(5)}\right)-\frac{G^2}{\omega_M}\left[\left(\hat{a}_1^{\dagger}\hat{a}_1 \right)^2 +\left(\hat{a}_2^{\dagger}\hat{a}_2 \right)^2 \right] \\\nonumber
               & +&\frac{\lambda_1\lambda_2}{\omega_M}\left( \sum_{i=4,5}\hat{a}_1\hat{a}^{\dagger}_2  \hat{\sigma}_{12}^{(i)}+h.c.\right)-\frac{G \lambda_1}{\omega_M}\left( \sum_{i=4,5}\hat{a}_1\hat{b}_1  \hat{\sigma}_{13}^{(i)}+h.c.\right)-\frac{G \lambda_2}{\omega_M}\left( \sum_{i=4,5}\hat{a}_2\hat{b}_2  \hat{\sigma}_{23}^{(i)}+h.c.\right),
               \end{eqnarray} 
          is performed between atoms (4,5). In fact, this effective Hamiltonian has been calculated using in a similar procedure as in Eq. (\ref{hamiltonian}) and the approach explained in Appendix A. The entangled state of optical and mechanical modes and atoms (1,4,5,8) may be obtained with the help of effective Hamiltonian (\ref{effectivehamiltonian2}) as below\footnote{Each ket of Eq. (\ref{state1-8}) is related to $\ket{(a_1, a^\dagger_1),(a_2, a^\dagger_2);(b_1, b^\dagger_1),(b_2, b^\dagger_2);\mathrm{atom}1,\mathrm{atom}4;\mathrm{atom}5,\mathrm{atom}8}$. The used method to calculate the coefficients of state (\ref{state1-8}) has been explained in Appendix C.}:
       \begin{eqnarray}\label{state1-8}
        \ket{\psi(\tau)}^i&=&B^i_1(\tau)\ket{0,0;0,0;\widetilde{1},\widetilde{3};\widetilde{3},\widetilde{1}}+B^i_2(\tau)\ket{0,0;0,0;\widetilde{1},\widetilde{3};\widetilde{1},\widetilde{3}}\\\nonumber
         &+&B^i_3(\tau)\ket{0,0;0,0;\widetilde{1},\widetilde{1};\widetilde{3},\widetilde{3}}+B^i_4(\tau)\ket{1,0;1,0;\widetilde{1},\widetilde{3};\widetilde{3},\widetilde{3}}\\\nonumber
          &+&B^i_5(\tau)\ket{0,0;0,0;\widetilde{3},\widetilde{1};\widetilde{1},\widetilde{3}}+B^i_6(\tau)\ket{1,0;1,0;\widetilde{3},\widetilde{3};\widetilde{1},\widetilde{3}}\\\nonumber
         &+&B^i_7(\tau)\ket{1,0;1,0;\widetilde{3},\widetilde{1};\widetilde{3},\widetilde{3}}+B^i_8(\tau)\ket{2,0;2,0;\widetilde{3},\widetilde{3};\widetilde{3},\widetilde{3}}\\\nonumber
          &+&B^i_9(\tau)\ket{0,0;0,0;\widetilde{3},\widetilde{1};\widetilde{3},\widetilde{1}}+B^i_{10}(\tau)\ket{0,0;0,0;\widetilde{3},\widetilde{3};\widetilde{1},\widetilde{1}}\\\nonumber
           &+&B^i_{11}(\tau)\ket{1,0;1,0;\widetilde{3},\widetilde{3};\widetilde{3},\widetilde{1}}.
       \end{eqnarray}
        In Eq. (\ref{state1-8}), $\tau$ is the time between atoms (4,5) and $i=1,2,3,4$ refers to the different initial states $\ket{0,0;0,0}\otimes\ket{\psi(t)}^{1}_{1,4}\otimes\ket{\psi(t)}^{1}_{5,8}$, $\ket{0,0;0,0}\otimes\ket{\psi(t)}^{2}_{1,4}\otimes\ket{\psi(t)}^{2}_{5,8}$, $\ket{0,0;0,0}\otimes\ket{\psi(t)}^{1}_{1,4}\otimes\ket{\psi(t)}^{2}_{5,8}$ and $\ket{0,0;0,0}\otimes\ket{\psi(t)}^{2}_{1,4}\otimes\ket{\psi(t)}^{1}_{5,8}$. Paying attention to Eqs. (\ref{state114}), (\ref{state214}), the initial state of Eq. (\ref{state1-8}) (when $\tau=t$) is considered as follows:
       \begin{eqnarray}\label{instate1-8}
        \ket{\psi(t)}^i&=&\ket{0,0;0,0}\otimes\left( \alpha(t)\ket{\widetilde{1},\widetilde{3};\widetilde{1},\widetilde{3}}+\beta(t)\ket{\widetilde{1},\widetilde{3};\widetilde{3},\widetilde{1}}+\gamma(t)\ket{\widetilde{3},\widetilde{1};\widetilde{1},\widetilde{3}}+\lambda(t)\ket{\widetilde{3},\widetilde{1};\widetilde{3},\widetilde{1}}\right),
       \end{eqnarray}
       where the coefficients of this state for $i=1,2,3,4$ are determined as below:
      \begin{itemize}
        \item{\textit{i}=1, \textit{i.e.,} when the initial state is $\ket{0,0;0,0}\otimes\ket{\psi(t)}^{1}_{1,4}\otimes\ket{\psi(t)}^{1}_{5,8}$, we have:}
       \begin{eqnarray}
        \alpha(t)=\dfrac{A^2_2(t)}{P(t)},\qquad  \beta(t)=\gamma(t)=\dfrac{A_2(t)A_{10}(t)}{P(t)},\qquad \lambda(t)=\dfrac{A^2_{10}(t)}{P(t)}.
       \end{eqnarray}
         \item{\textit{i}=2, \textit{i.e.,} when the initial state is $\ket{0,0;0,0}\otimes\ket{\psi(t)}^{2}_{1,4}\otimes\ket{\psi(t)}^{2}_{5,8}$}, we have:
         \begin{eqnarray}
          \alpha(t)=\dfrac{A^2_{10}(t)}{P(t)},\qquad  \beta(t)=\gamma(t)=\dfrac{A_2(t)A_{10}(t)}{P(t)},\qquad \lambda(t)=\dfrac{A^2_{2}(t)}{P(t)}.
         \end{eqnarray}
         \item{\textit{i}=3, \textit{i.e.,} when the initial state is $\ket{0,0;0,0}\otimes\ket{\psi(t)}^{1}_{1,4}\otimes\ket{\psi(t)}^{2}_{5,8}$}, we have:
         \begin{eqnarray}
          \alpha(t)=\lambda(t)=\dfrac{A_2(t)A_{10}(t)}{P(t)},\qquad  \beta(t)=\dfrac{A^2_2(t)}{P(t)},\qquad \gamma(t)=\dfrac{A^2_{10}(t)}{P(t)}.
         \end{eqnarray}
           \item{\textit{i}=4, \textit{i.e.,} when the initial state is $\ket{0,0;0,0}\otimes\ket{\psi(t)}^{2}_{1,4}\otimes\ket{\psi(t)}^{1}_{5,8}$, we have:}
             \begin{eqnarray}
              \alpha(t)=\lambda(t)=\dfrac{A_2(t)A_{10}(t)}{P(t)},\qquad  \beta(t)=\dfrac{A^2_{10}(t)}{P(t)},\qquad \gamma(t)=\dfrac{A^2_{2}(t)}{P(t)}.
             \end{eqnarray}
       \end{itemize}
       Now, by applying the projection operator which is performed with the state $\ket{0,0} \otimes \ket{\widetilde{3},\widetilde{1}}$ \footnote{Note that the ket $|0, 0 \rangle$ is related to optical and mechanical mode denoted by $(a_1, a^\dagger_1)$ and $(b_1, b^\dagger_1)$, respectively, also, the state $|\widetilde{3},\widetilde{1}\rangle$ is related to atoms 4, 5.} on state (\ref{state1-8}), the atoms (1,8) are converted to the following entangled state,
       \begin{eqnarray}\label{state18}
           \ket{\psi(\tau)}^i_{(1,8)}&=&\frac{1}{\sqrt{P^i_{1,8}(\tau)} }(B^i_2(\tau)\ket{\widetilde{1},\widetilde{3}}+B^i_{10}(\tau)\ket{\widetilde{3},\widetilde{1}}),
        \end{eqnarray}
         with the corresponding entropy and success probability as
           \begin{eqnarray}\label{ent18}
             E^{i}_{1,8}(\tau)&=&1-\frac{\left| B^i_2(\tau)\right| ^4+\left| B^i_{10}(\tau)\right|^4}{(P^i_{1,8}(\tau))^2},
               \end{eqnarray}
         \begin{eqnarray}\label{suc18}
        P^{i}_{1,8}(\tau)&=&\left| B^i_2(\tau)\right| ^2+\left| B^i_{10}(\tau)\right|^2.
          \end{eqnarray}
        Similarly, doing the above-mentioned procedure by the state $\ket{0,0} \otimes \ket{\widetilde{1},\widetilde{3}}$ \footnote{Note that the ket $|0, 0 \rangle$ is related to optical and mechanical mode denoted by $(a_1, a^\dagger_1)$ and $(b_1, b^\dagger_1)$, respectively, also, the state $|\widetilde{1},\widetilde{3}\rangle$ is related to atoms 4, 5.}, the atoms (1,8) are transformed to the following entangled state,
       \begin{eqnarray}\label{state218}
       \ket{\psi^{'}(\tau)}^i_{(1,8)}&=&\frac{1}{\sqrt{P^{'i}_{1,8}(\tau)} }(B^i_3(\tau)\ket{\widetilde{1},\widetilde{3}}+B^i_9(\tau)\ket{\widetilde{3},\widetilde{1}}),
        \end{eqnarray}
        with the associated entropy and success probability as
          \begin{eqnarray}\label{ent218}
            E^{'i}_{1,8}(\tau)&=&1-\frac{\left| B^i_3(\tau)\right| ^4+\left| B^i_9(\tau)\right|^4}{(P^{'i}_{1,8}(\tau))^2},
              \end{eqnarray}
        \begin{eqnarray}\label{suc218}
        P^{'i}_{1,8}(\tau)&=&\left| B^i_3(\tau)\right| ^2+\left| B^i_9(\tau)\right|^2.
          \end{eqnarray}
        Now we are ready to present our numerical results about the entropies and success probabilities, the task that may be followed in the next section.
   \section{Results and discussion} \label{sec.results}
  In this section we consider the effects of mechanical frequency, $\omega_M$, and optomechanical coupling strength to the field modes, $G$, on the evolution of entropy and success probability related to atoms (1,4) (or (5,8)) and specially to atoms (1,8). In figures \ref{fig.fig2} and \ref{fig.fig3}, the effects of $\omega_M$ and $G$ on the evolution of entropy and success probability related to atoms (1,4) (or (5,8)) have been considered\footnote{To find differences between the results achieved in this paper and our previous work \cite{Ghasemi42019}, the amounts of $\omega_M$ and $G$ in this paper have been selected similar to the parameters introduced in Ref. \cite{Ghasemi42019}, so the presented figures \ref{fig.Fig2a}, \ref{fig.Fig2b} and \ref{fig.fig3} are similar to figures 2, 3 in Ref. \cite{Ghasemi42019}.}. As is observed from figures \ref{fig.Fig2a}, \ref{fig.Fig2b}, \ref{fig.Fig2c} and \ref{fig.Fig2d}, the time periods of entropy and success probability are increased by increasing $\omega_M$. But, in figures \ref{fig.Fig3a} and \ref{fig.Fig3b} the time periods of entropy and success probability are decreased by increasing $G$. Also, entanglement sudden death and its revival is observed in figures \ref{fig.Fig2a}, \ref{fig.Fig2c} and \ref{fig.Fig3a}. The effects of $G$ on the evolution of entropy and success probability related to target atoms (1,8), after initial interaction time $\lambda_1 t=0.8$ (interaction time between atoms (2,3) and (6,7)) have been shown in figure \ref{fig.fig4}. In figures \ref{fig.Fig4a}, \ref{fig.Fig4c} and \ref{fig.Fig4e} the time periods of entropy are decreased by increasing $G$. Also, in figures \ref{fig.Fig4b} and \ref{fig.Fig4d}, the maximum of success probability is increased by decreasing $G$, while, in figure \ref{fig.Fig4f} the success probability is increased by increasing $G$.\\
   In figure \ref{fig.fig5} we consider the effect of $\omega_M$ on the evolution of entropy and success probability related to target atoms (1,8). As is clear from figures \ref{fig.Fig5a}, \ref{fig.Fig5c} and \ref{fig.Fig5e} the maximum entanglement of target atomic state is more accessible by decreasing $\omega_M$. In figures \ref{fig.Fig5b} and \ref{fig.Fig5d}, the maximum of success probability has been increased by decreasing $\omega_M$, but in figure \ref{fig.Fig5f}, the success probability has been increased by increasing $\omega_M$.
      \begin{figure}[H]
                 \centering
                 \subfigure[\label{fig.Fig2a} \ $E(t)$]{\includegraphics[width=0.45\textwidth]{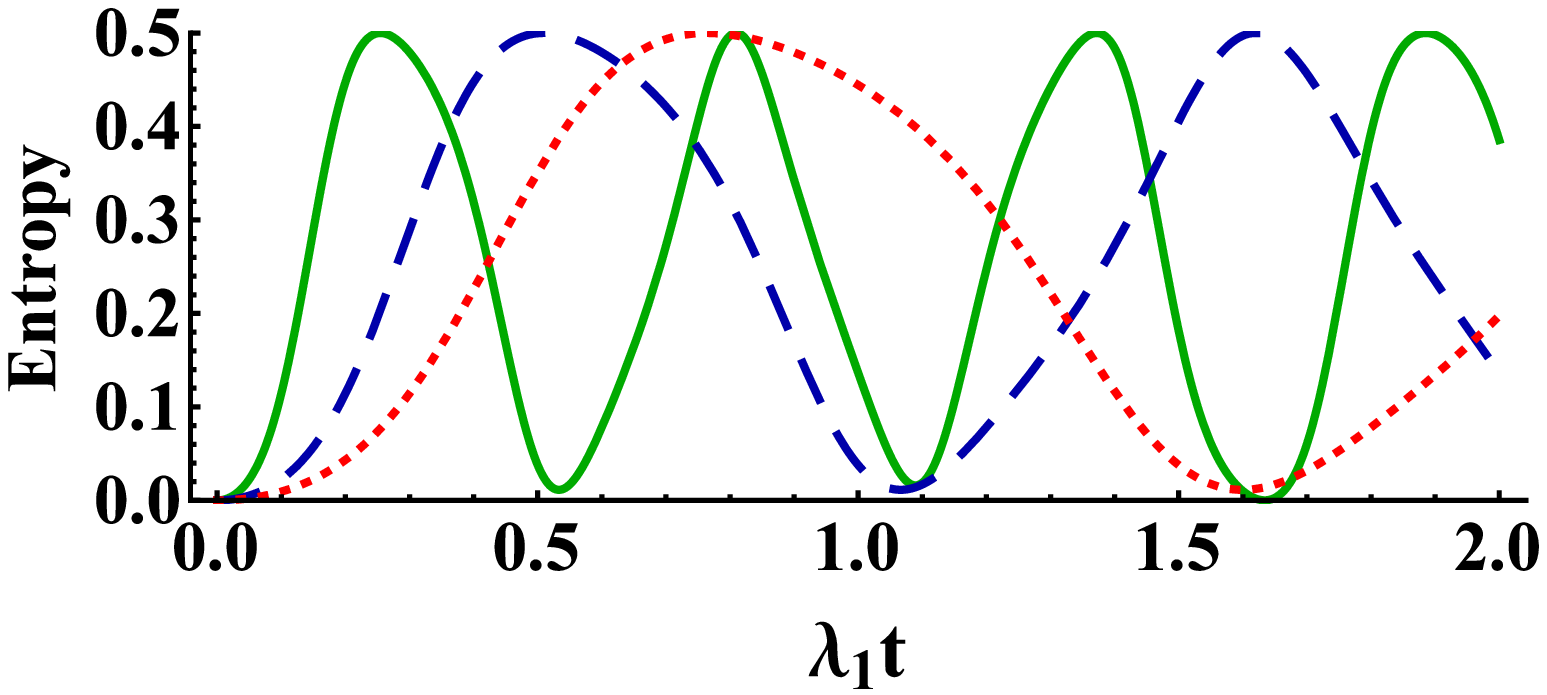}}
                 \hspace{0.05\textwidth}
                 \subfigure[\label{fig.Fig2b} \ $P(t)$]{\includegraphics[width=0.45\textwidth]{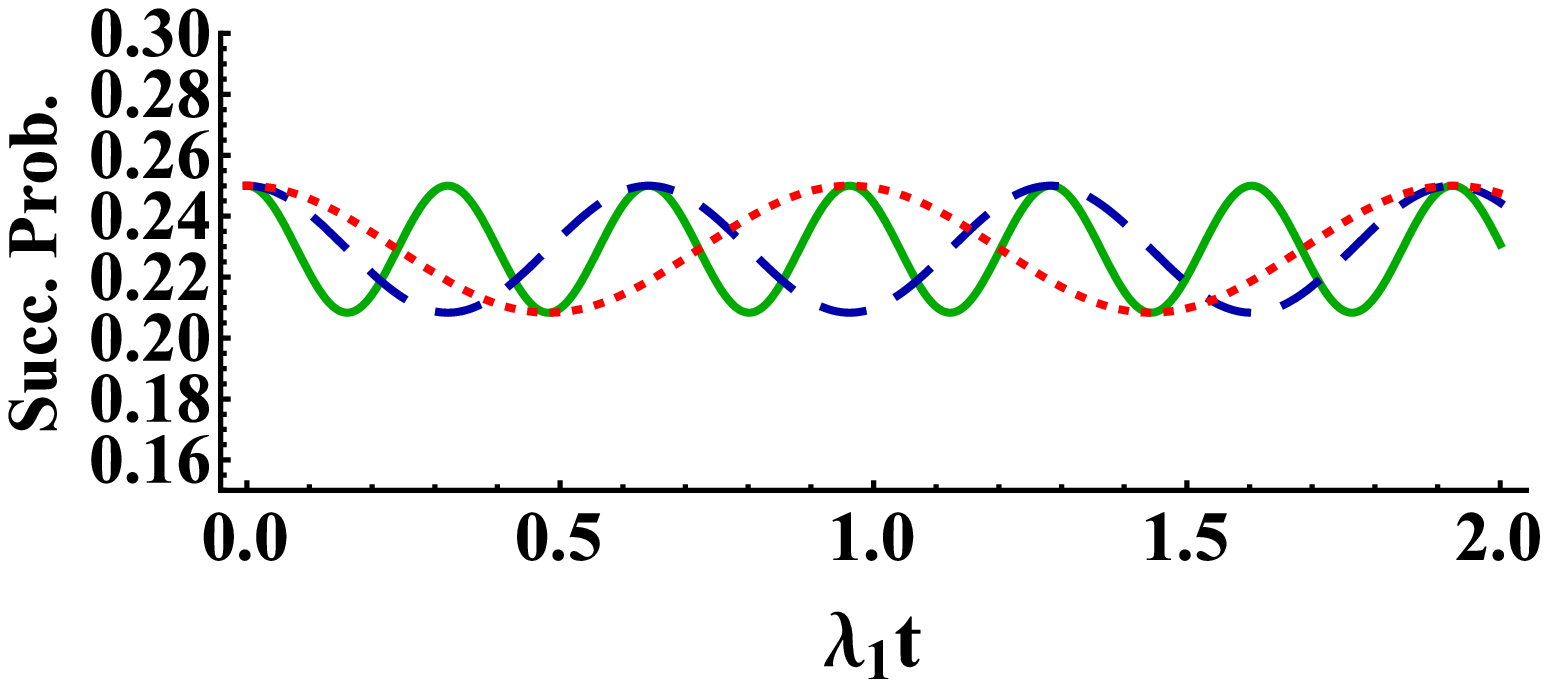}}
                  \hspace{0.05\textwidth}
                 \subfigure[\label{fig.Fig2c} \ $E(t)$]{\includegraphics[width=0.45\textwidth]{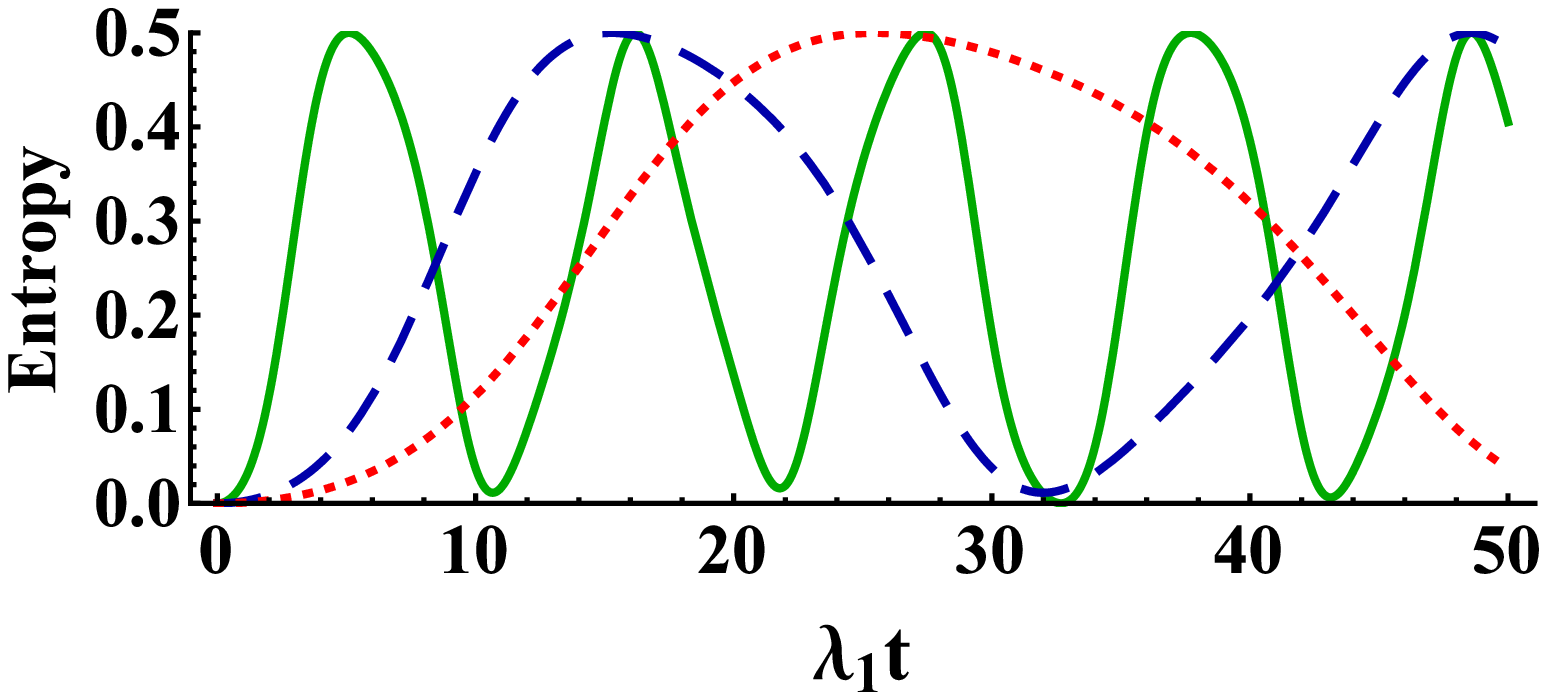}}
                \hspace{0.05\textwidth}
               \subfigure[\label{fig.Fig2d} \ $P(t)$]{\includegraphics[width=0.45\textwidth]{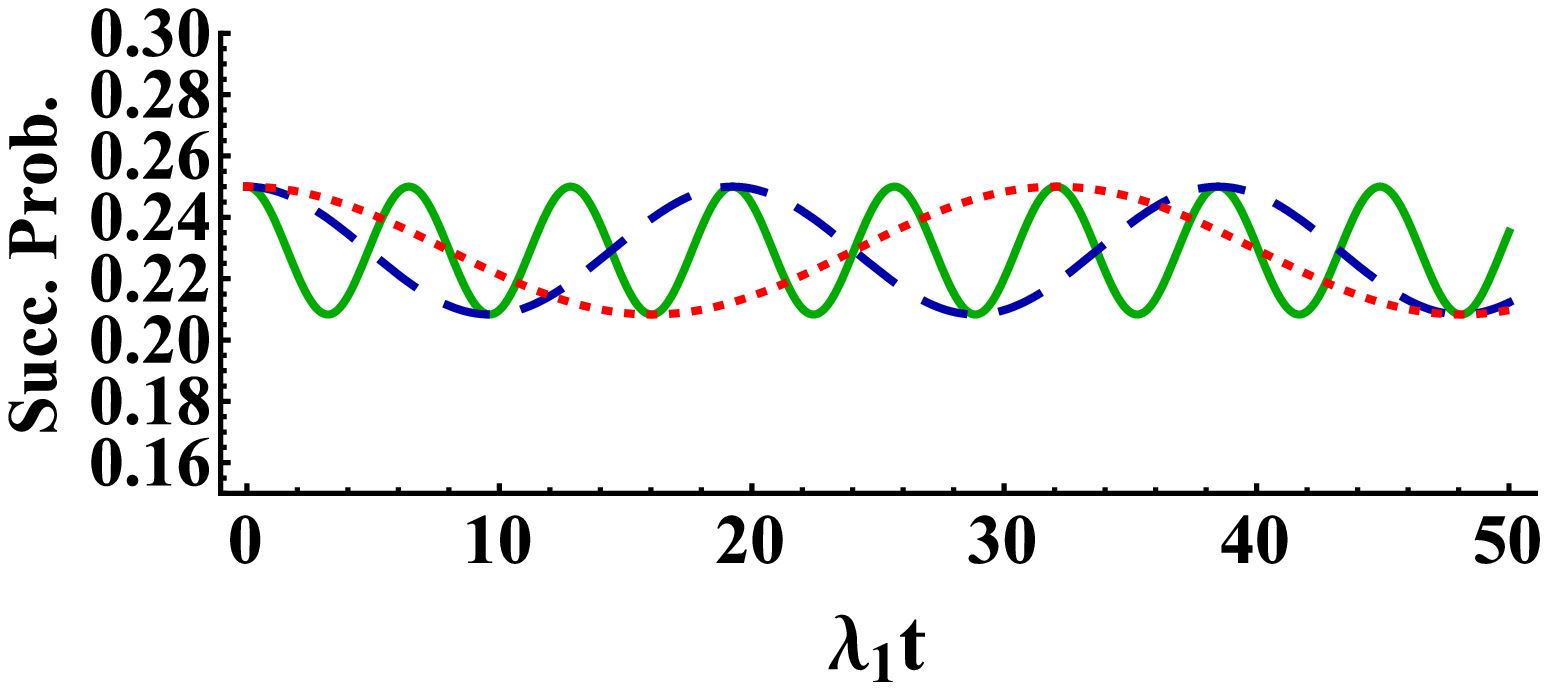}}
                  \caption{\label{fig.fig2} {\it The effect of mechanical frequency, $\omega_M$, on the evolution of} (a) entropy (Eq. (\ref{ent14})), (b) success probability (Eq. (\ref{suc14})) for $\omega_M/\lambda_1=0.5$ (solid green line), $\omega_M/\lambda_1=1$ (dashed blue line), $\omega_M/\lambda_1=1.5$ (dotted red line), (c) entropy (Eq. (\ref{ent14})) and (d) success probability (Eq. (\ref{suc14})) for $\omega_M/\lambda_1=10$ (solid green line), $\omega_M/\lambda_1=30$ (dashed blue line) and  $\omega_M/\lambda_1=50$ (dotted red line) with $G/\lambda_1=2$.}
                 \end{figure}
           \begin{figure}[H]
            \centering
            \subfigure[\label{fig.Fig3a} \ $E(t)$]{\includegraphics[width=0.45\textwidth]{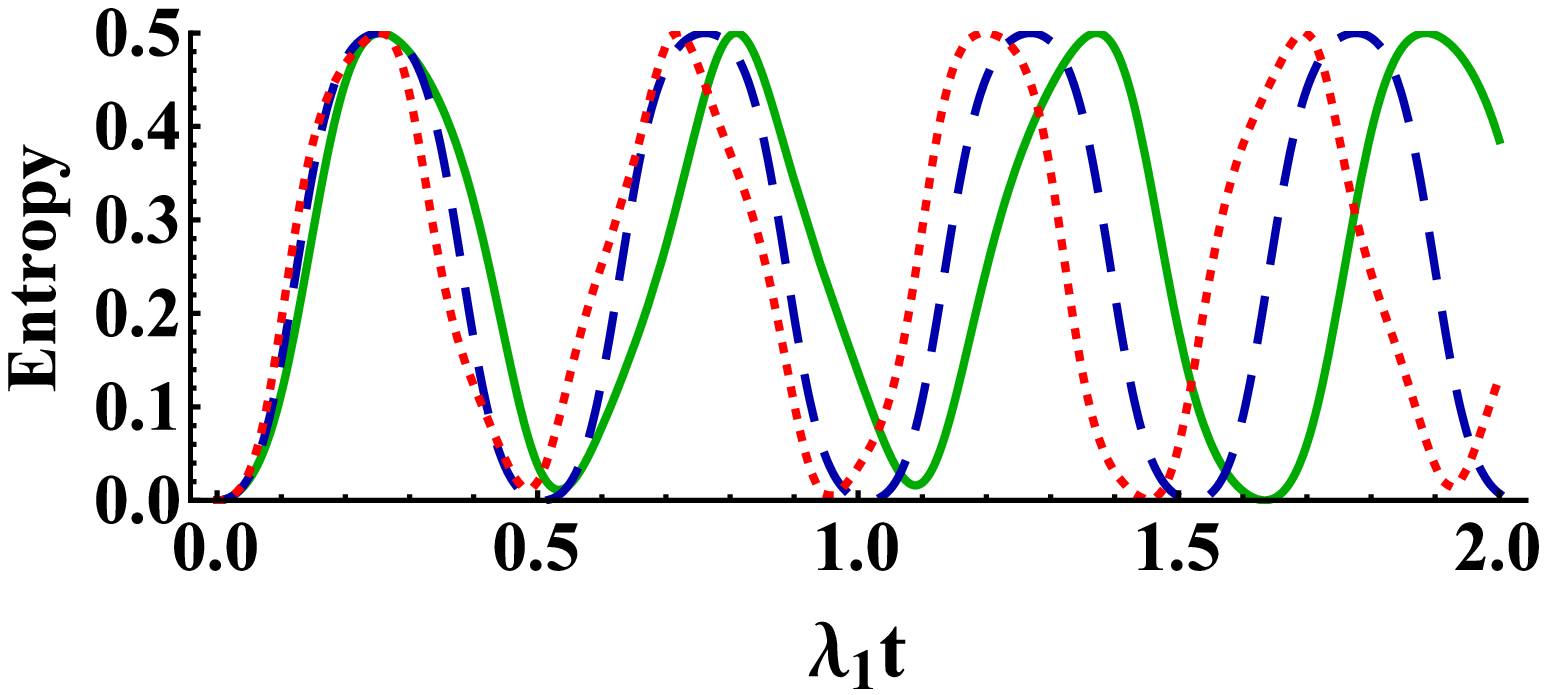}}
            \hspace{0.05\textwidth}
            \subfigure[\label{fig.Fig3b} \  $P(t)$]{\includegraphics[width=0.45\textwidth]{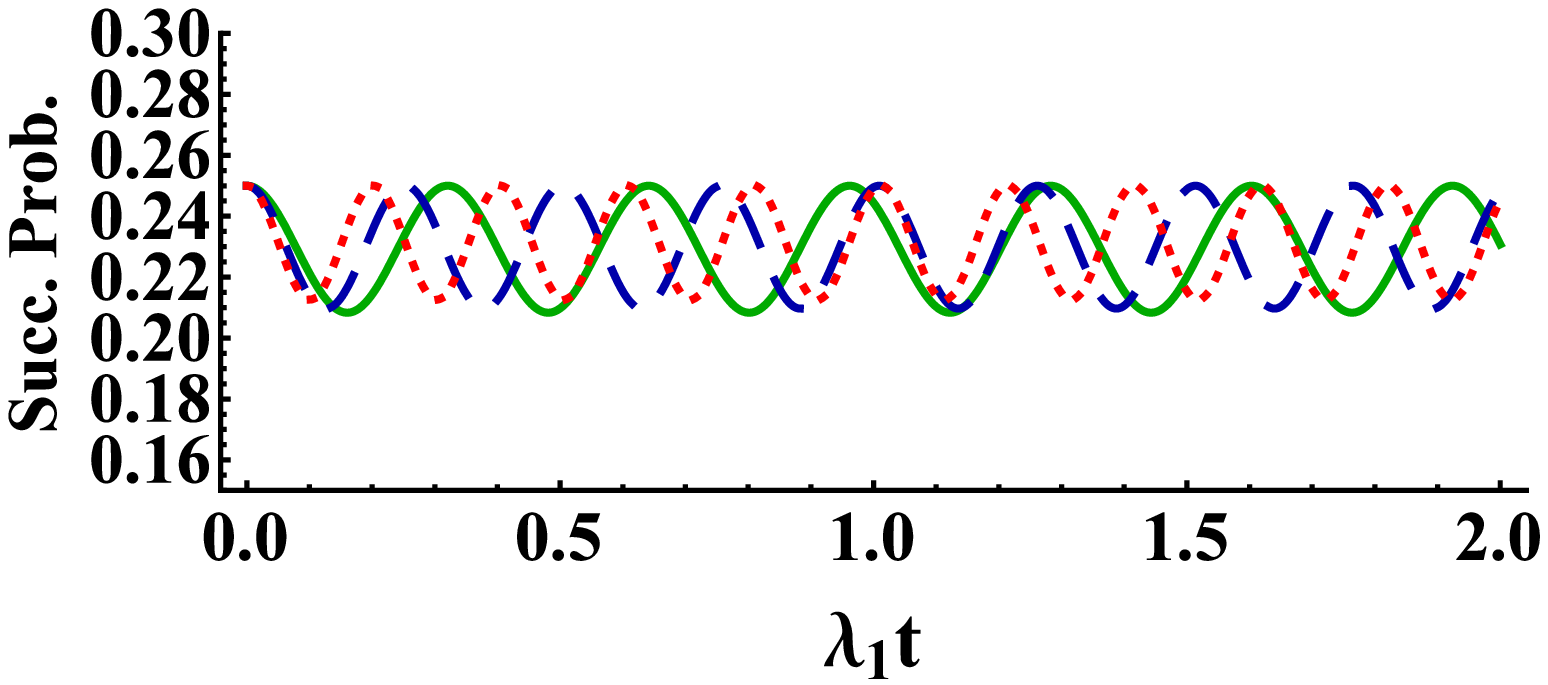}}
           \caption{\label{fig.fig3} {\it The effect of optomechanical coupling strength to the field modes, $G$, on the evolution of} (a) entropy (Eq. (\ref{ent14})) and (b) success probability (Eq. (\ref{suc14})) for $G/\lambda_1=2$ (solid green line), $G/\lambda_1=2.5$ (dashed blue line), $G/\lambda_1=3$ (dotted red line) with $\omega_M/\lambda_1=0.5$.}
            \end{figure}
         \begin{figure}[H]
            \centering
            \subfigure[\label{fig.Fig4a} \ $E^1_{1,8}(\tau)=E^{'2}_{1,8}(\tau)$]{\includegraphics[width=0.45\textwidth]{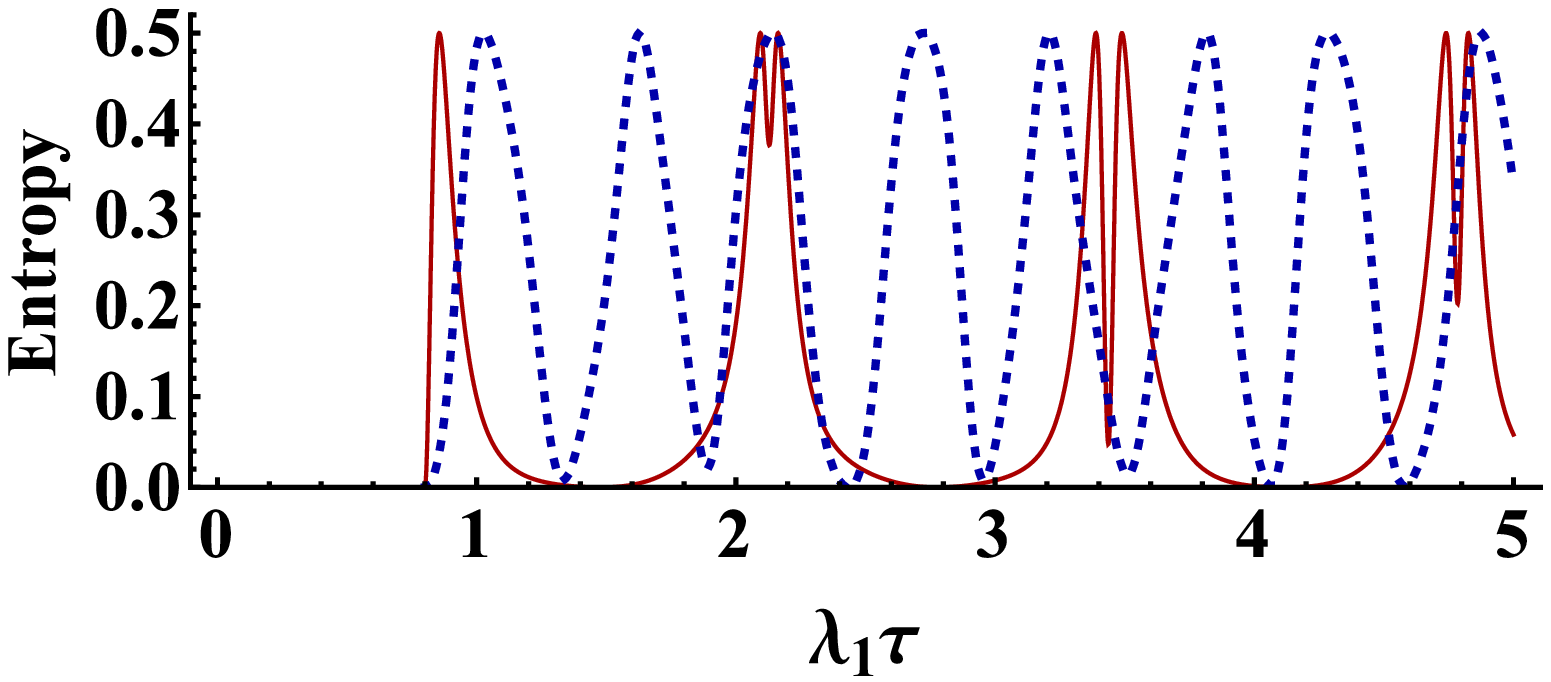}}
            \hspace{0.05\textwidth}
            \subfigure[\label{fig.Fig4b} \  $P^1_{1,8}(\tau)=P^{'2}_{1,8}(\tau)$]{\includegraphics[width=0.45\textwidth]{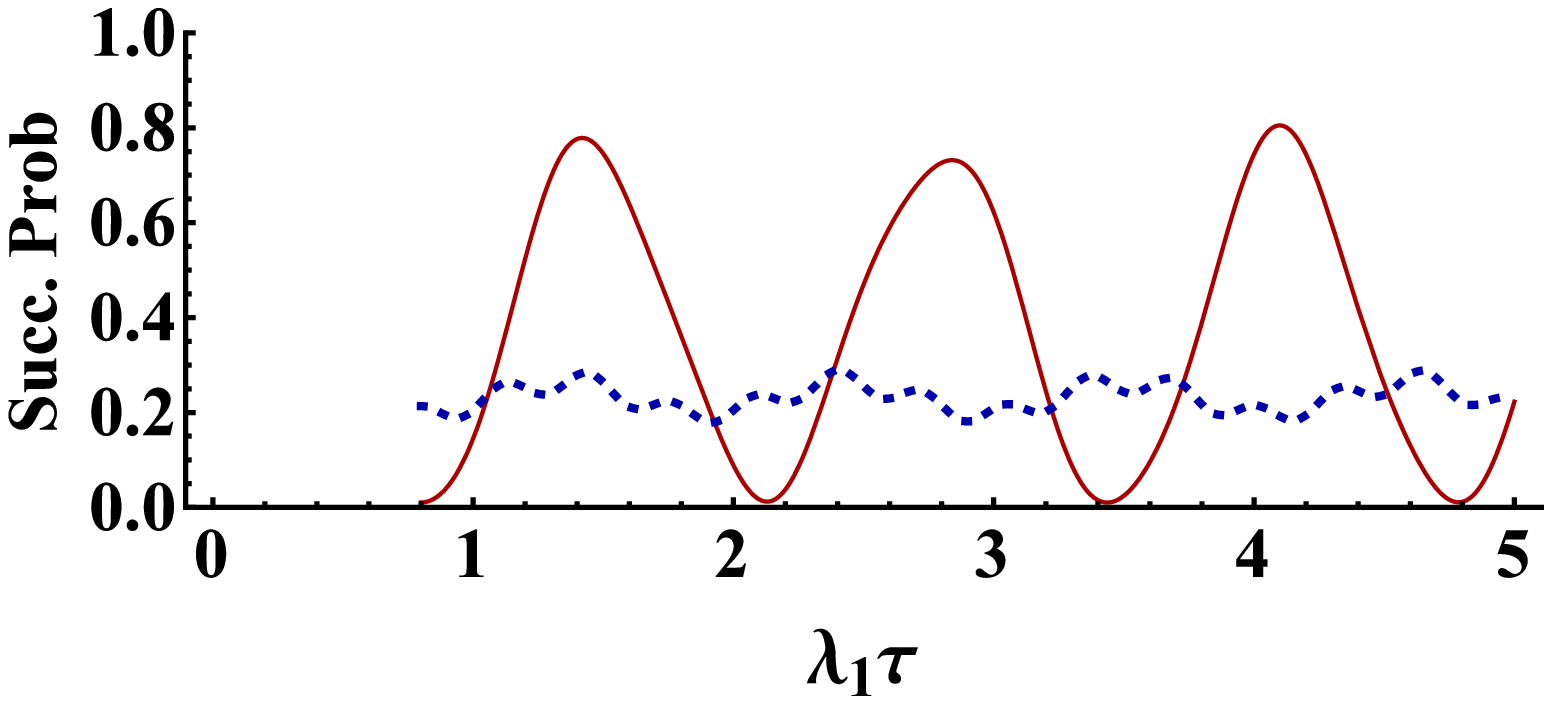}}
             \hspace{0.05\textwidth}
                \subfigure[\label{fig.Fig4c} \  $E^2_{1,8}(\tau)=E^{'1}_{1,8}(\tau)$]{\includegraphics[width=0.45\textwidth]{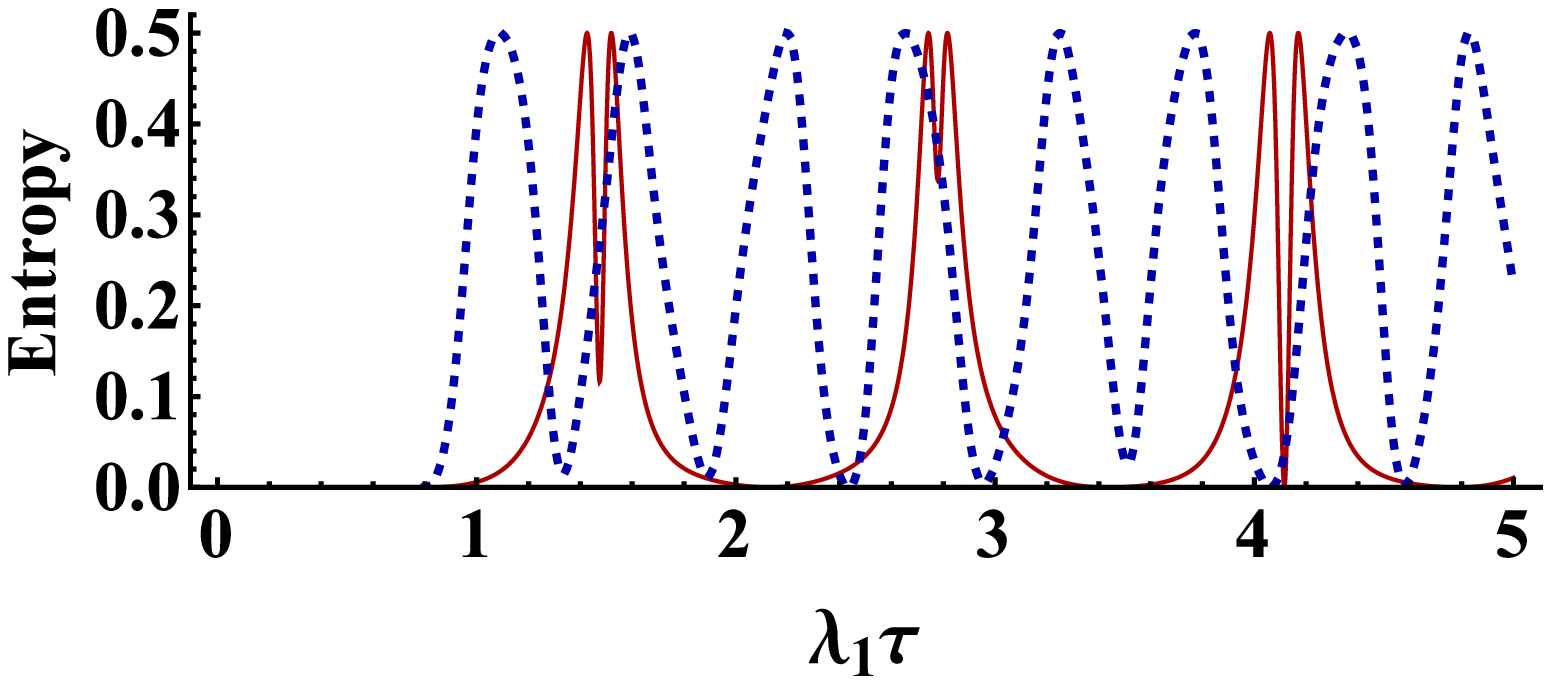}}
                 \hspace{0.05\textwidth}
                    \subfigure[\label{fig.Fig4d} \  $P^2_{1,8}(\tau)=P^{'1}_{1,8}(\tau)$]{\includegraphics[width=0.45\textwidth]{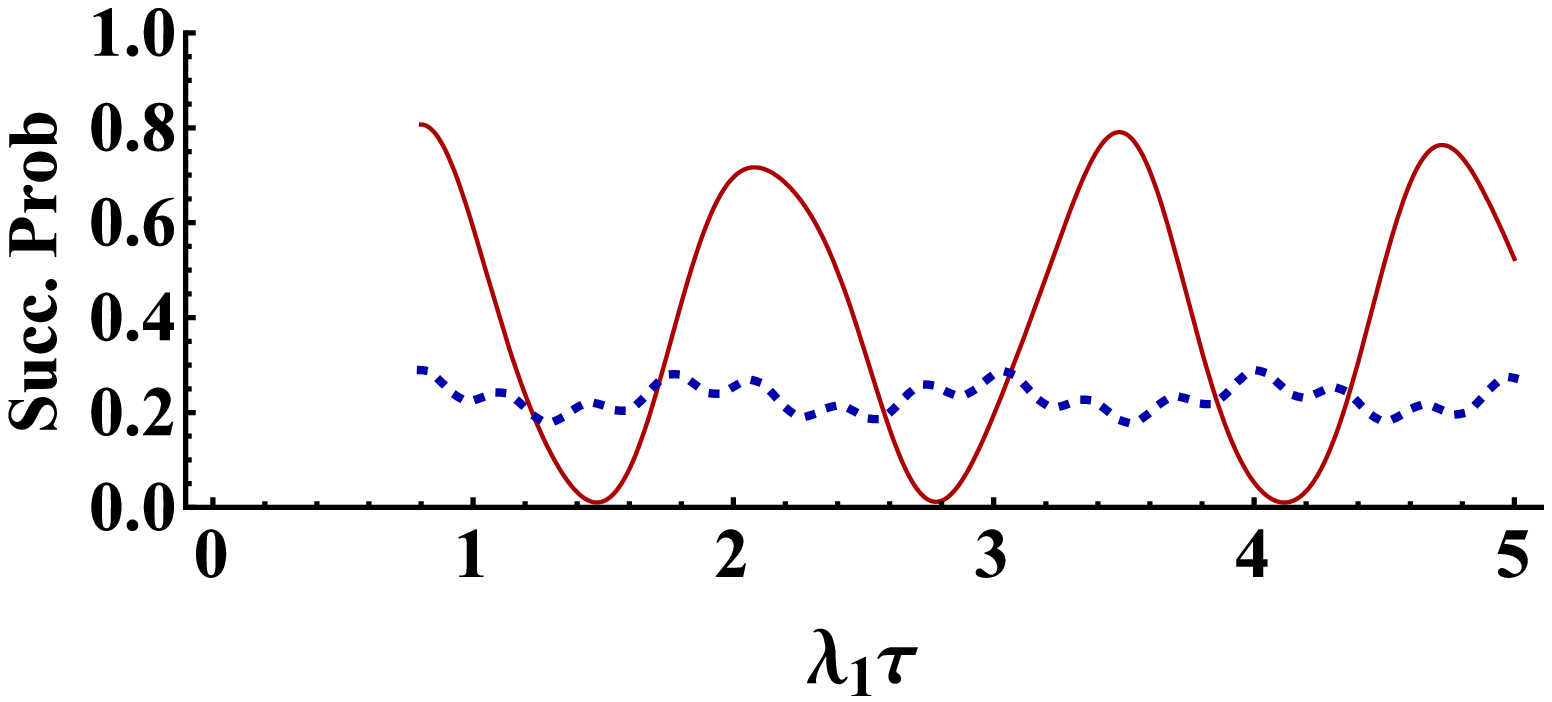}}
                     \hspace{0.05\textwidth}
                        \subfigure[\label{fig.Fig4e} \   $E^3_{1,8}(\tau)=E^4_{1,8}(\tau)=E^{'3}_{1,8}(\tau)=E^{'4}_{1,8}(\tau)$]{\includegraphics[width=0.45\textwidth]{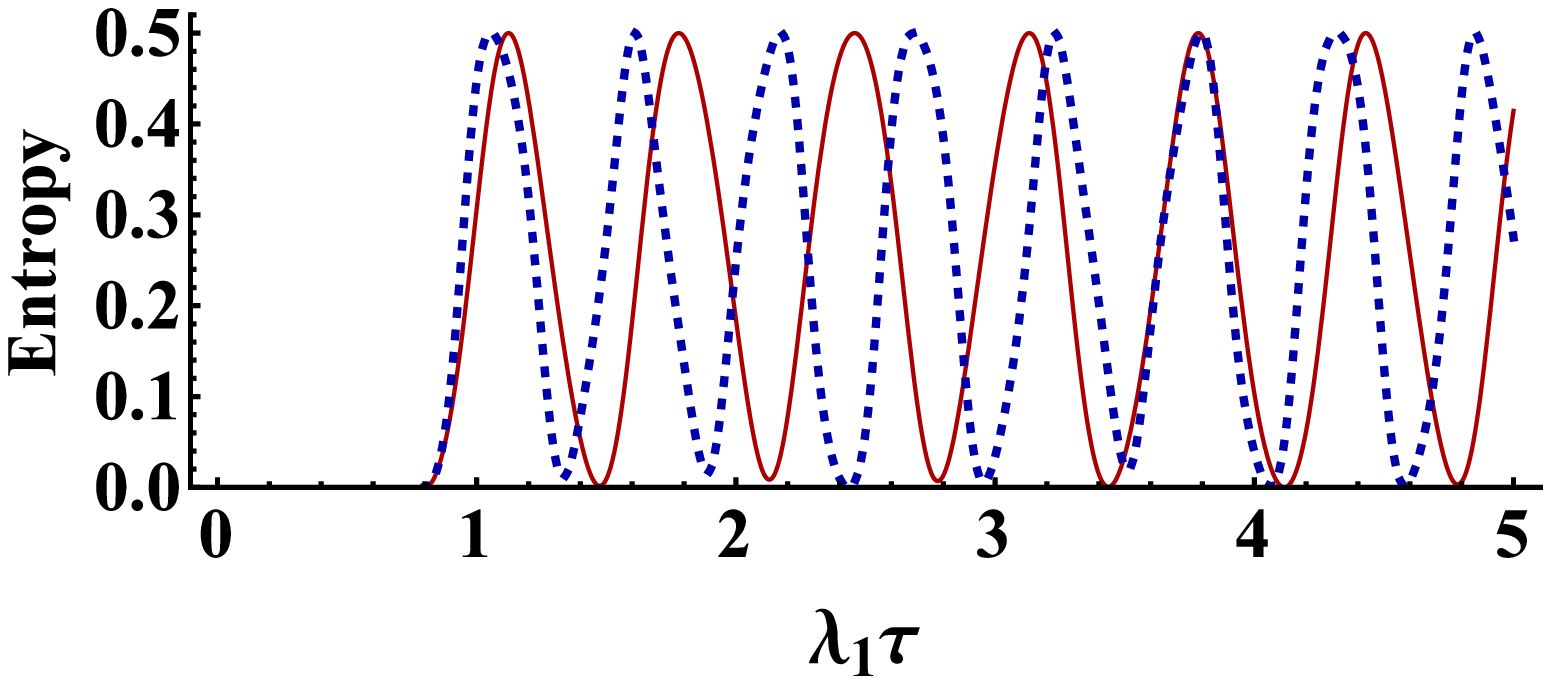}}
                         \hspace{0.05\textwidth}
                            \subfigure[\label{fig.Fig4f} \  $P^3_{1,8}(\tau)=P^4_{1,8}(\tau)=P^{'3}_{1,8}(\tau)=P^{'4}_{1,8}(\tau)$]{\includegraphics[width=0.45\textwidth]{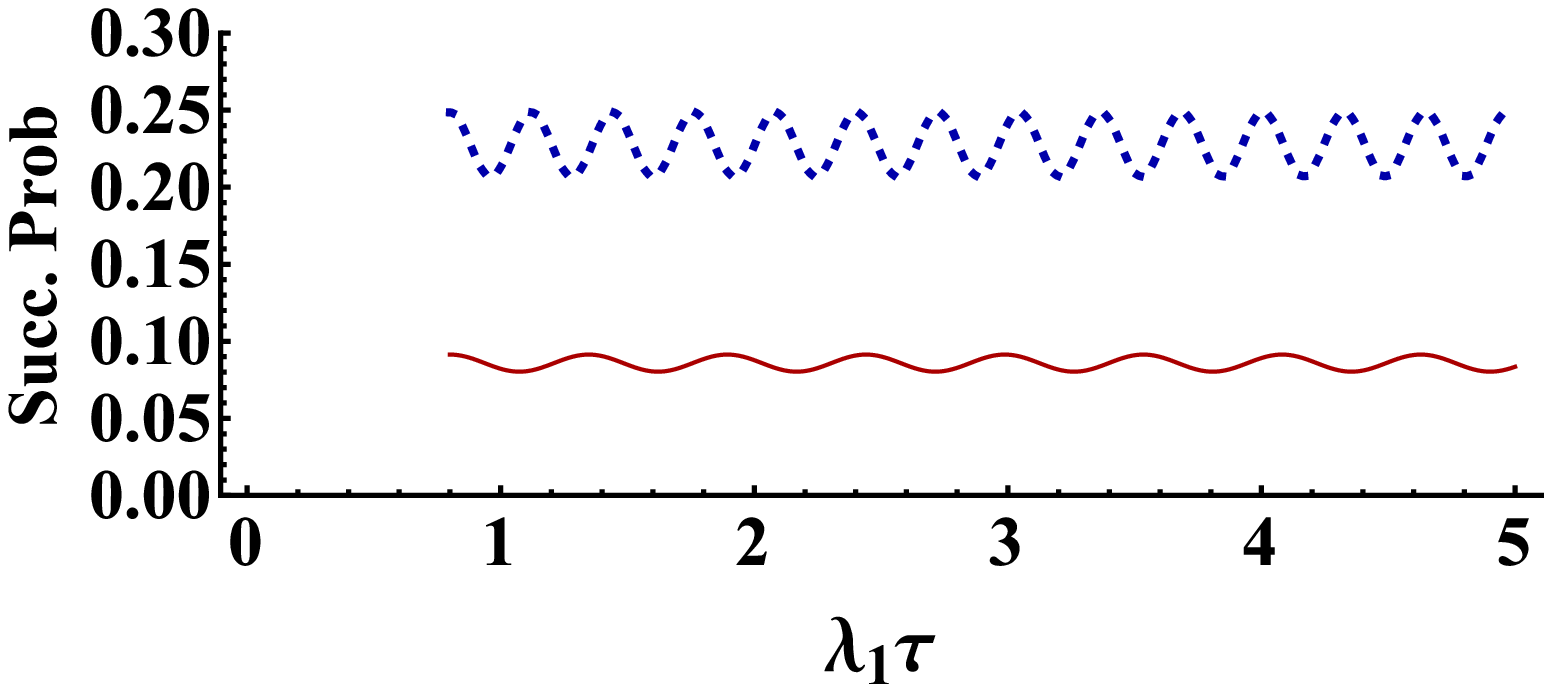}}
            \caption{\label{fig.fig4} {\it The effect of optomechanical coupling strength to the field modes, $G$, on the evolution of} (a) entropy (Eqs. (\ref{ent18}), (\ref{ent218})) and (b) success probability (Eqs. (\ref{suc18}), (\ref{suc218})) for $G/\lambda_1=1$ (solid red line), $G/\lambda_1=2$ (dotted blue line) with $\omega_M/\lambda_1=0.5$ and $\lambda_1 t=0.8$.}
            \end{figure}
         \begin{figure}[H]
            \centering
            \subfigure[\label{fig.Fig5a} \ $E^1_{1,8}(\tau)=E^{'2}_{1,8}(\tau)$]{\includegraphics[width=0.45\textwidth]{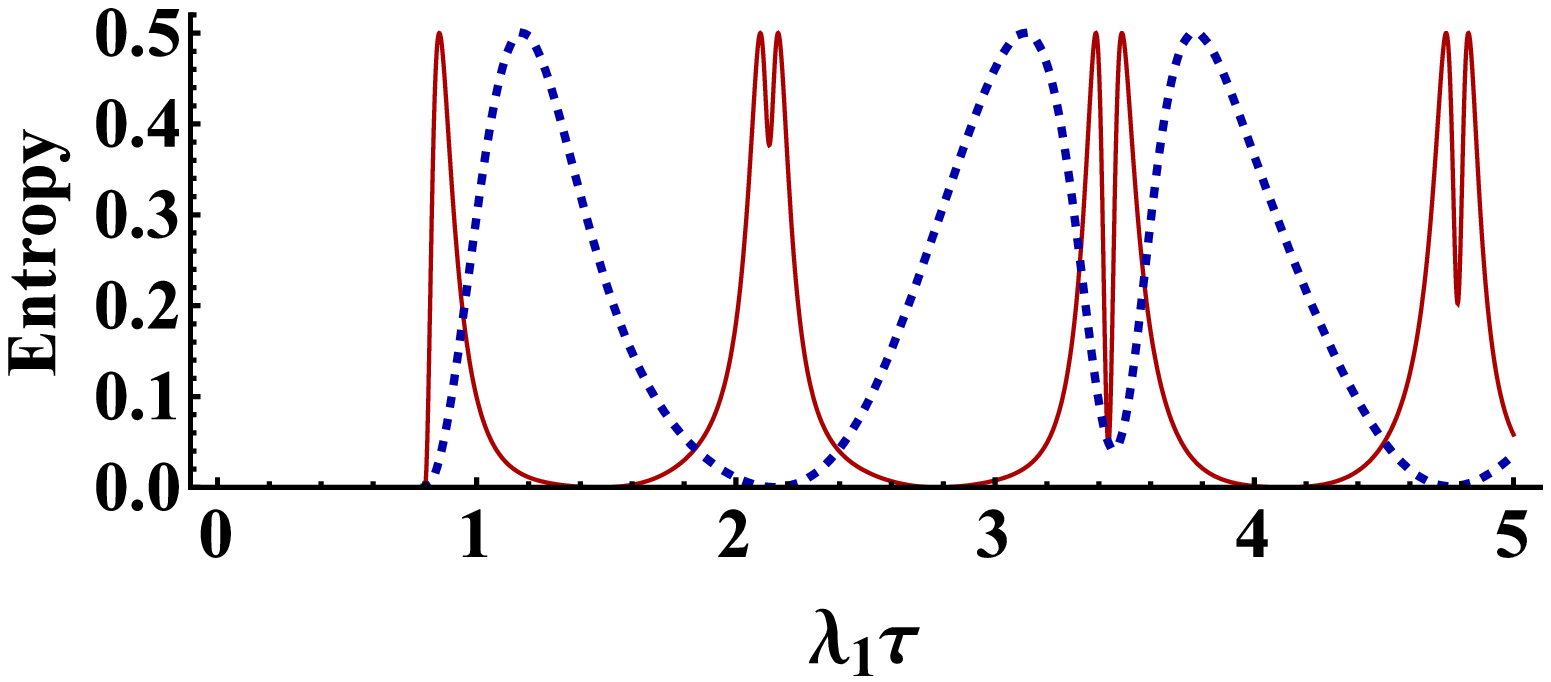}}
            \hspace{0.05\textwidth}
            \subfigure[\label{fig.Fig5b} \  $P^1_{1,8}(\tau)=P^{'2}_{1,8}(\tau)$]{\includegraphics[width=0.45\textwidth]{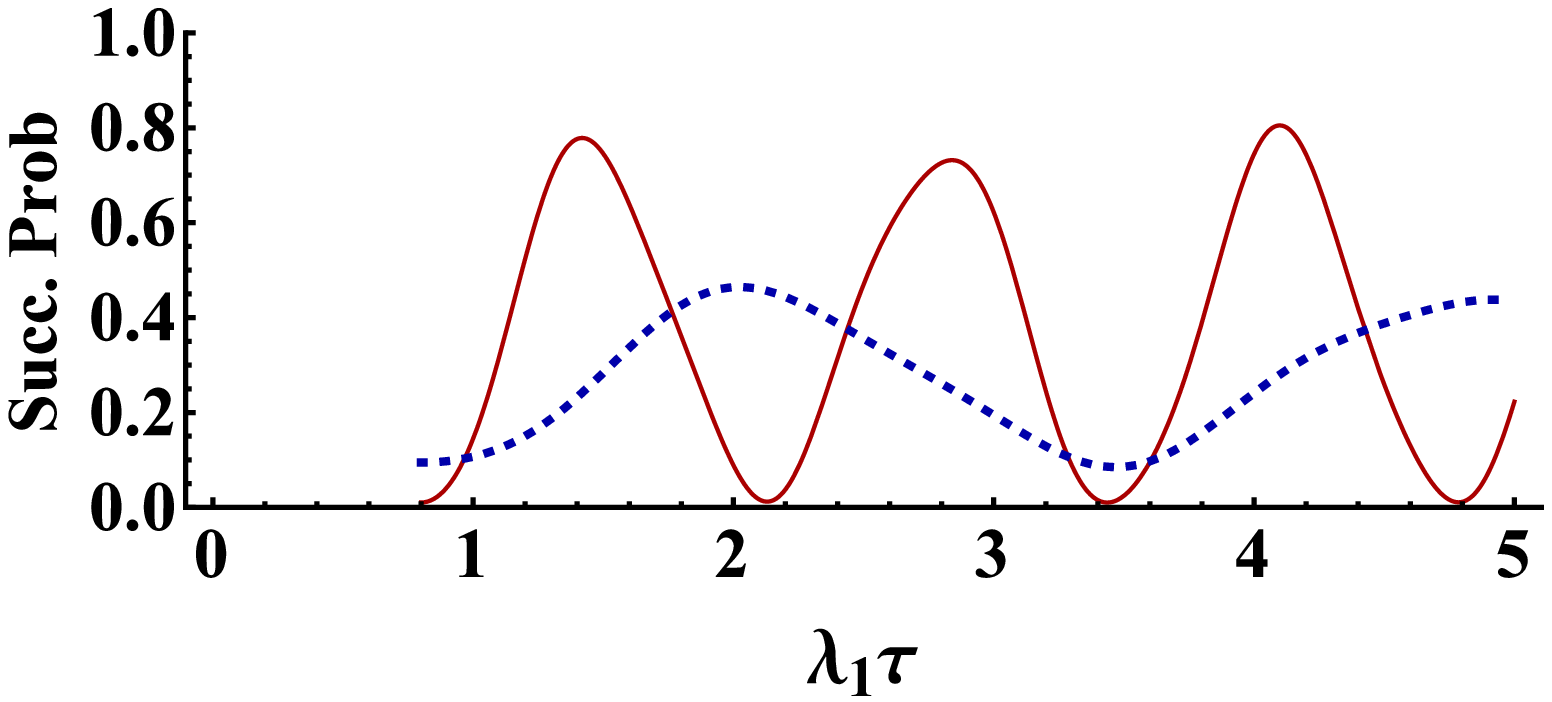}}
             \hspace{0.05\textwidth}
                \subfigure[\label{fig.Fig5c} \  $E^2_{1,8}(\tau)=E^{'1}_{1,8}(\tau)$]{\includegraphics[width=0.45\textwidth]{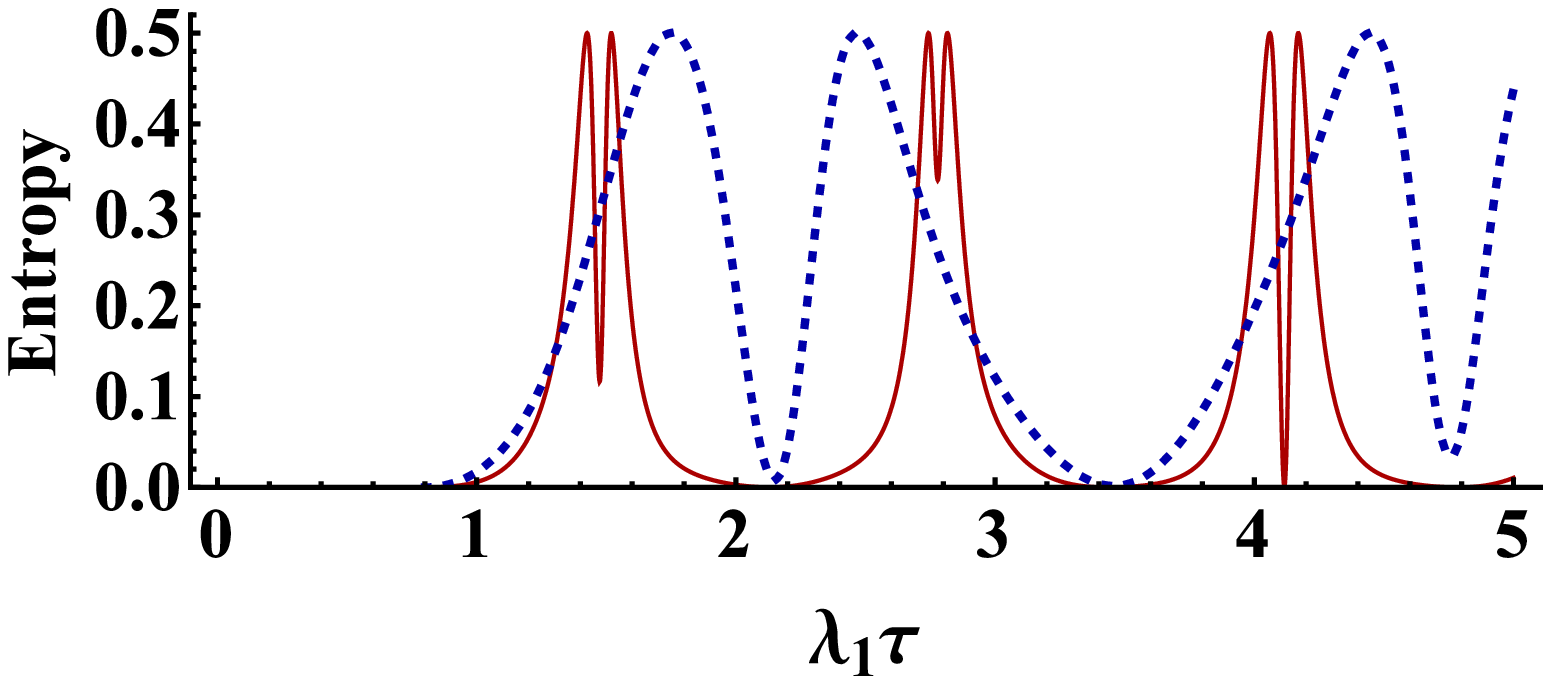}}
                 \hspace{0.05\textwidth}
                    \subfigure[\label{fig.Fig5d} \  $P^2_{1,8}(\tau)=P^{'1}_{1,8}(\tau)$]{\includegraphics[width=0.45\textwidth]{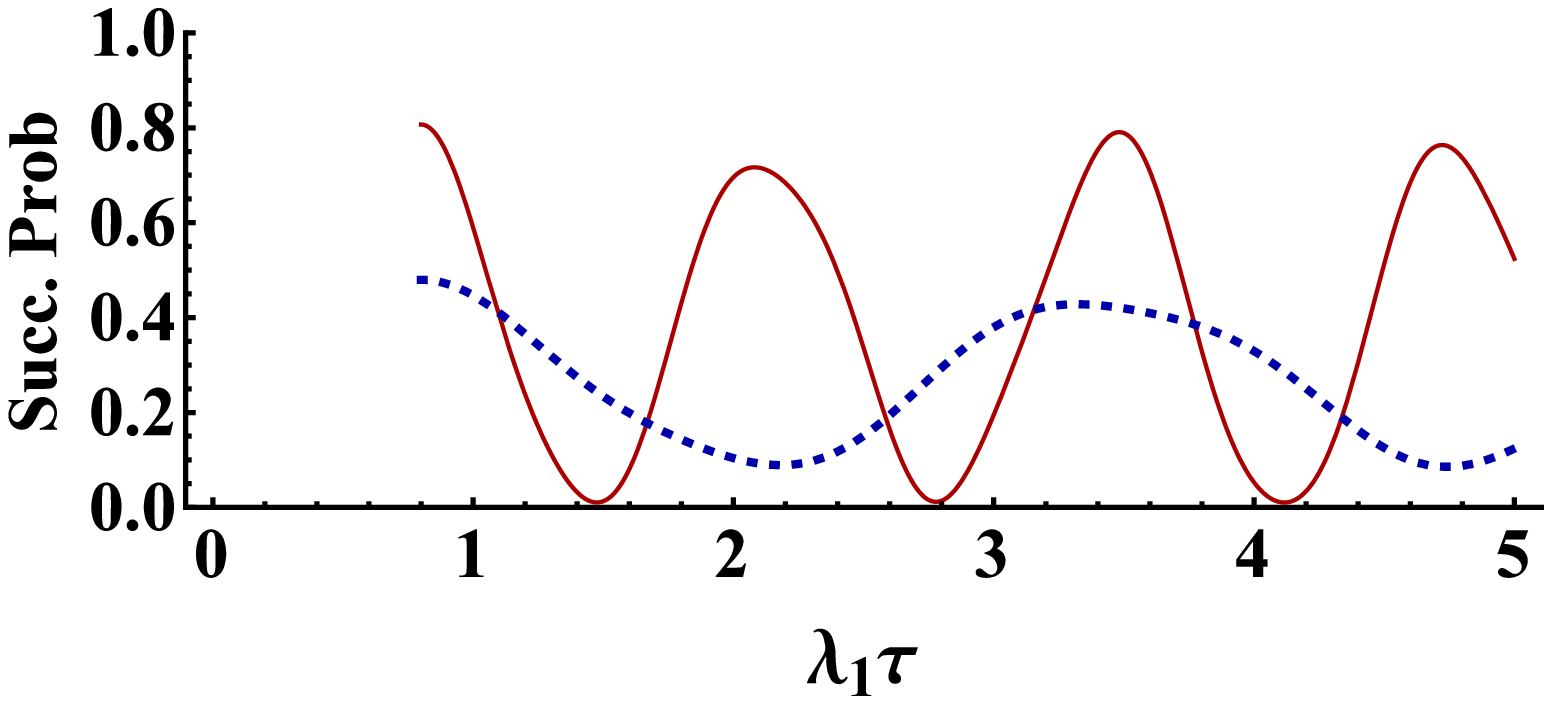}}
                     \hspace{0.05\textwidth}
                        \subfigure[\label{fig.Fig5e} \   $E^3_{1,8}(\tau)=E^4_{1,8}(\tau)=E^{'3}_{1,8}(\tau)=E^{'4}_{1,8}(\tau)$]{\includegraphics[width=0.45\textwidth]{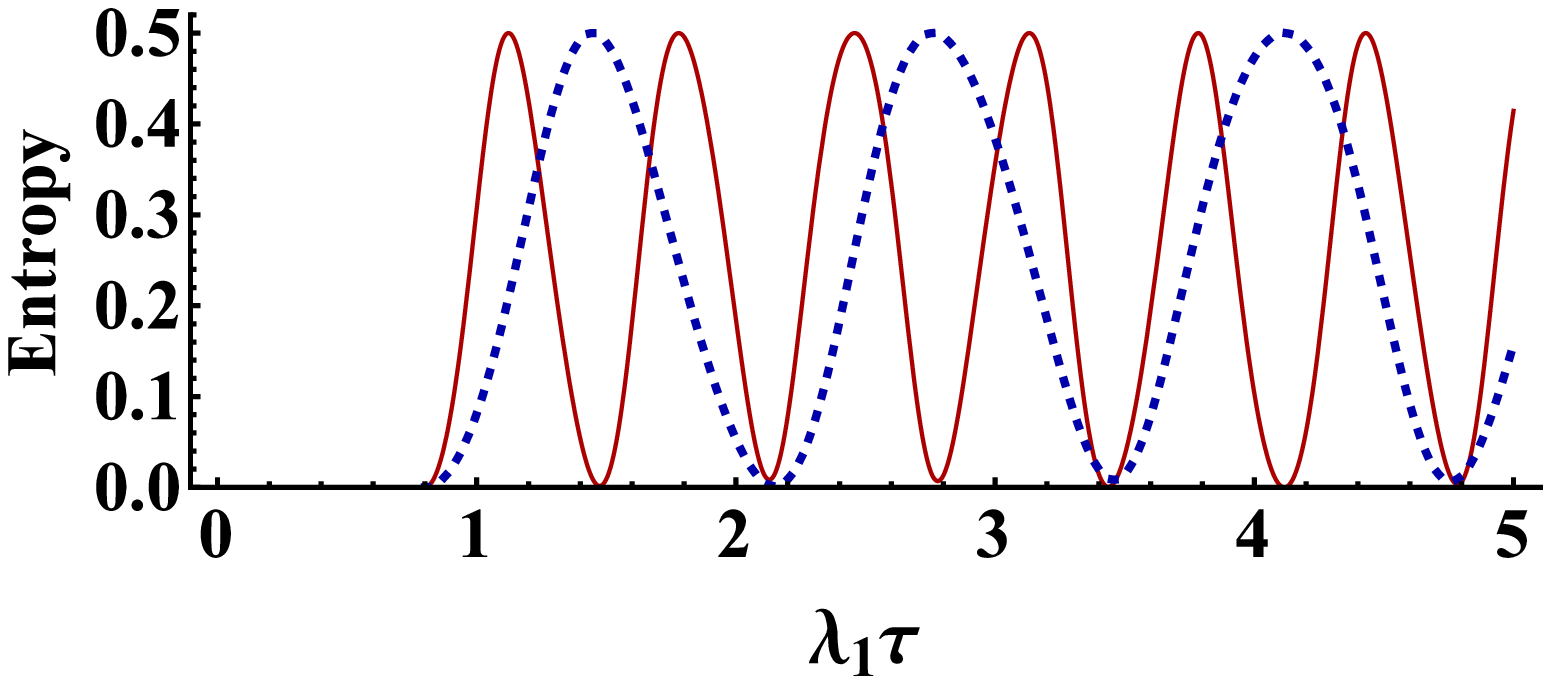}}
                         \hspace{0.05\textwidth}
                            \subfigure[\label{fig.Fig5f} \  $P^3_{1,8}(\tau)=P^4_{1,8}(\tau)=P^{'3}_{1,8}(\tau)=P^{'4}_{1,8}(\tau)$]{\includegraphics[width=0.45\textwidth]{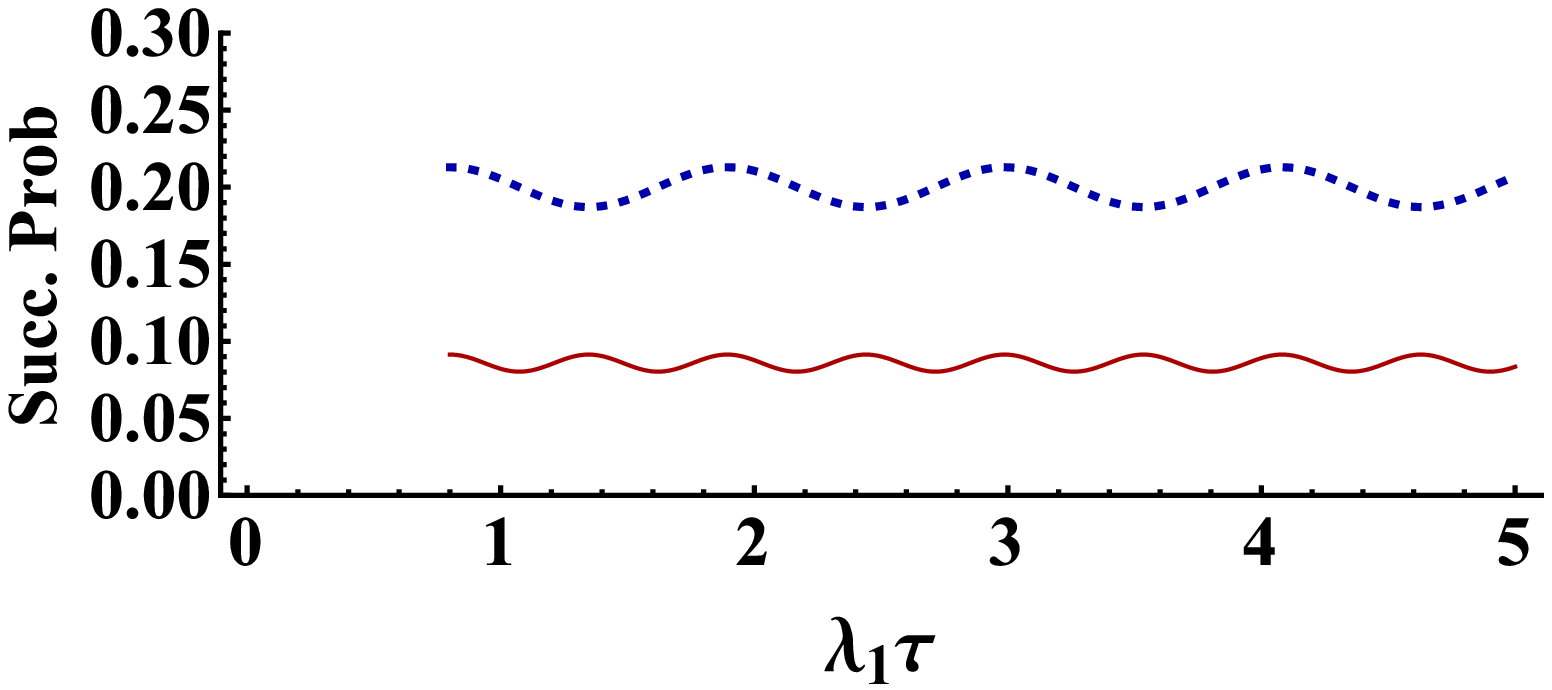}}
           \caption{\label{fig.fig5} {\it The effect of mechanical frequency, $\omega_M$, on the evolution of} (a) entropy (Eqs. (\ref{ent18}), (\ref{ent218})) and (b) success probability (Eqs. (\ref{suc18}), (\ref{suc218})) for $\omega_M/\lambda_1=0.5$ (solid red line), $\omega_M/\lambda_1=1$ (dotted blue line) with $G/\lambda_1=1$ and $\lambda_1 t=0.8$.}
            \end{figure}
         \section{Summary and conclusions}\label{Summary}
  In this paper, we produced distributed entangled states using quantum repeater protocol in an arrangement that has been fully constructed by OMCs. In this protocol, we considered eight three-level $V$-type atoms numbered by $(1,2 \cdots 8)$ where the pairs $(i,i+1)$ with $i=1,3,5,7$ have been initially prepared in some entangled states, while especially the end atoms, \textit{i.e.,} atoms 1, 8 are the target separable far atoms which should be entangled, as our purpose.  Then, by performing interaction between atoms (2,3) and (6,7) in two OMCs and operating proper measurement, the pairs (1,4) and (5,8) were converted into entangled states, respectively. Finally, the state of target atoms (1,8) was transformed to an entangled state after performing interaction between atoms (4,5) in another OMC and operating appropriate  measurement. Note that we have not used the Bell state measurement in any case. Our numerical results show that the entanglement and success probability of the quantum repeater process can be tuned via the OMC parameters, \textit{i.e.,} optomechanical coupling strength to the field modes $G$ and mechanical frequency $\omega_M$. As is observed, the time periods of entropy of the produced entangled state were increased by decreasing $G$ and increasing $\omega_M$; also, the maximum of success probability can be increased by decreasing $G$ and $\omega_M$. Therefore, by preparing a weak coupling between mechanical-optical (phonon-photon) modes we can achieve an acceptable entanglement and success probability between the far atoms (1, 8).

  \section{Appendix A: The used approach to obtain the effective Hamiltonian (\ref{effectivehamiltonian})}\label{Appeff}
  As mentioned prior to Eq. (\ref{effectivehamiltonian}), following the approach explained in Refs. \cite{James2007,Gamel2010}, the effective Hamiltonian (\ref{effectivehamiltonian}) can be obtained. In this approach, at first using the Hamiltonian (\ref{hamiltonian}) and the following Baker-Hausdorff lemma:
     \begin{eqnarray}\label{intH}
       \hat{H}^\mathrm{int}(t)&=&e^{i \hat{H}_0 t} \hat{H}_1e^{-i \hat{H}_0 t}\\ \nonumber
       &=&\hat{H}_1+it\left[\hat{H}_0,\hat{H}_1 \right]+\frac{(it)^2}{2!} \left[\hat{H}_0,\left[\hat{H}_0,\hat{H}_1 \right]\right]+\cdots
        \end{eqnarray}
  the Hamiltonian in the interaction picture is obtained. As stated in Refs. \cite{James2007,Gamel2010}, after calculating the Hamiltonian in the interaction picture with the following form:
             \begin{eqnarray}\label{int}
             \hat{H}^{\mathrm{int}}(t)=\sum_{n=1}^{N}\hat{h}_n \exp({-i\omega_{n}t})+\hat{h}_n^{\dagger} \exp({i\omega_{n}t}),
             \end{eqnarray}
             the effective Hamiltonian can be obtained as below:
            \begin{eqnarray}\label{eff2}
            \hat{H}^{\mathrm{eff}}(t)=\sum_{m,n=1}^{N}{1\over\hbar{\omega}_{{mn}}}[\hat{h}_m^{\dagger},\hat{h}_n] \exp\left({i\left[\omega_{m}-\omega_{n}\right]}t\right).
            \end{eqnarray}
      In Eqs. (\ref{int}), (\ref{eff2}), $\omega_{n(m)}>0$ is
                           frequency and $N$ is the total number of different harmonic terms
                     making up the interaction Hamiltonian. Also, ${\omega}_{{mn}}$ is defined by the following equation:
                     \begin{eqnarray}\label{A13}
                     {1\over{\omega}_{{mn}}}={1\over2}\left({1\over\omega_m}+{1\over\omega_n}\right),
                     \end{eqnarray}
                    which may be considered as the harmonic average of $\omega_{m}$ and $\omega_{n}$.   
     In our case, the effective Hamiltonian (\ref{effectivehamiltonian}) has been obtained using Eqs. (\ref{hbl}), (\ref{int}), (\ref{eff2}) and (\ref{A13}).
  \section{Appendix B: Calculating the coefficients of entangled state (\ref{state1-4})}\label{App} 
   In this section, the differential equations related to the state (\ref{state1-4}) with the help of effective Hamiltonian (\ref{effectivehamiltonian}) and the time-dependent Schr\"{o}dinger equation are obtained as $\dot{X}=SX$, where $\dot{X}$ and $S$ are respectively defined as follow,
         \begin{footnotesize}
          \begin{eqnarray}
          \scriptsize
             \dot{X}&=& \dfrac{d}{dt} \left( \begin{array}{ccccccccccc}
              A_1(t) &  A_2(t) & A_3(t) & A_4(t) & A_5(t) & A_6(t) & A_7(t) & A_8(t) & A_9(t) & A_{10}(t) & A_{11}(t)  \\
             \end{array} \right)^T,\\\nonumber
             \label{eq:unitarymatrix1}
           \end{eqnarray}
           \end{footnotesize}
            \begin{equation}
            \tiny
              S= \left( \begin{array}{ccccccccccc}
               0  & 0 & 0 & 0 & 0 & 0 & 0 & 0 & 0 & 0 & 0\\
               0  & -i\dfrac{\lambda^2_1}{\omega_M} & -i\dfrac{\lambda^2_1}{\omega_M} & i\dfrac{G \lambda_1}{\omega_M} & 0 & 0 & 0 & 0 & 0 & 0 & 0\\
               0  & -i\dfrac{\lambda^2_1}{\omega_M} & -i\dfrac{\lambda^2_1}{\omega_M} & i\dfrac{G \lambda_1}{\omega_M} & 0 & 0 & 0 & 0 & 0 & 0 & 0 \\
               0  &  i\dfrac{G \lambda_1}{\omega_M} &  i\dfrac{G \lambda_1}{\omega_M} &  i\dfrac{ (2\lambda^2_1+ G^2)}{\omega_M} & 0 & 0 & 0 & 0 & 0 & 0 & 0  \\
               0  &  0 &  0 & 0 &  -2 i\dfrac{ \lambda^2_1}{\omega_M} & i\dfrac{G \lambda_1}{\omega_M} & i\dfrac{G \lambda_1}{\omega_M} & 0 & 0 & 0 & 0  \\
               0  &  0 &  0 & 0 & i\dfrac{G \lambda_1}{\omega_M} & -i\dfrac{ (\lambda^2_1-G^2)}{\omega_M} & -i\dfrac{\lambda^2_1}{\omega_M} & 2 i\dfrac{ G \lambda_1}{\omega_M} & 0 & 0 & 0  \\
               0  &  0 &  0 & 0 & i\dfrac{G \lambda_1}{\omega_M} & -i\dfrac{\lambda^2_1}{\omega_M} &  -i\dfrac{ (\lambda^2_1-G^2)}{\omega_M} & 2 i\dfrac{G \lambda_1}{\omega_M} & 0 & 0 & 0  \\
               0  &  0 &  0 & 0 & 0 & 2 i\dfrac{G \lambda_1}{\omega_M} &  2 i\dfrac{G \lambda_1}{\omega_M} & 4 i\dfrac{ (\lambda^2_1+ G^2)}{\omega_M} & 0 & 0 & 0  \\
               0  & 0 & 0 & 0 & 0 & 0 & 0 & 0 & -i\dfrac{\lambda^2_1}{\omega_M} & -i\dfrac{\lambda^2_1}{\omega_M} & i\dfrac{G \lambda_1}{\omega_M}\\
               0  & 0 & 0 & 0 & 0 & 0 & 0 & 0 & -i\dfrac{\lambda^2_1}{\omega_M} & -i\dfrac{\lambda^2_1}{\omega_M} & i\dfrac{G \lambda_1}{\omega_M}\\
               0  & 0 & 0 & 0 & 0 & 0 & 0 & 0 & i\dfrac{G \lambda_1}{\omega_M} & i\dfrac{G \lambda_1}{\omega_M} & i\dfrac{(2 \lambda^2_1+ G^2)}{\omega_M}\\ \end{array} \right).
             \label{eq:unitarymatrix2}
             \end{equation}
     The equation $\dot{A}_1(t)=0$ easily results in $A_1(t)=\frac{1}{2}$. Also, paying attention to the considered initial conditions, we find that $A_2(t)=A_9(t)$, $A_3(t)=A_{10}(t)$, $A_4(t)=A_{11}(t)$ and $A_6(t)=A_{7}(t)$. Now, using Laplace transform and with the help of Mathematica software, the coefficients $A_4(t)$, $A_5(t)$, $A_8(t)$ and then the coefficients $A_2(t)$, $A_3(t)$ and $A_6(t)$ can be calculated, however, due to their complex and lengthy forms we ignore presenting their explicit forms.
      \section{Appendix C: Calculating the coefficients of entangled state (\ref{state1-8})}\label{App2} 
       The differential equations related to state (\ref{state1-8}) using the effective Hamiltonian (\ref{effectivehamiltonian2}) and the time-dependent Schr\"{o}dinger equation are obtained as $\dot{Y}=SY$, where $S$ has been defined in (\ref{eq:unitarymatrix2}) and $\dot{Y}$ is defined as follows,
             \begin{footnotesize}
              \begin{eqnarray}
              \scriptsize
                 \dot{Y}&=& \dfrac{d}{d\tau} \left( \begin{array}{ccccccccccc}
                          B^i_1(\tau) & B^i_2(\tau) & B^i_3(\tau) & B^i_4(\tau) & B^i_5(\tau) & B^i_6(\tau) & B^i_7(\tau) & B^i_8(\tau) & B^i_9(\tau) & B^i_{10}(\tau) & B^i_{11}(\tau)  \\
                         \end{array} \right)^T.
               \label{eq:unitarymatrix3}
               \end{eqnarray}
               \end{footnotesize}
Paying attention to Eqs. (\ref{state18}), (\ref{state218}), it may be seen that, knowing the coefficients $B^i_{2}(\tau)$, $B^i_{3}(\tau)$, $B^i_{9}(\tau)$ and $B^i_{10}(\tau)$ is enough for our purpose, and these coefficients can be obtained after calculating $B^i_{4}(\tau)$ and $B^i_{11}(\tau)$ using Laplace transform and with the help of Mathematica software.


\end{document}